\documentclass[12pt]{article}
\usepackage{amscd}
\usepackage{verbatim}
\usepackage{hyperref}
\usepackage{amssymb}
\usepackage{amsmath}
\usepackage{tabularx}
\usepackage{graphicx}
\usepackage{mathtools}

\DeclarePairedDelimiterX\braket[2]{\langle}{\rangle}{#1 \delimsize\vert #2}
\newcommand{\RomanNumeralCaps}[1]
    {\MakeUppercase{\romannumeral #1}}

\graphicspath{ {/User/sam/Documents/Latex/Second law 9 may} }
\begin{document}
\begin{center}
\vspace{24pt} { \large \bf A local Generalized second law in crossed product constructions} \\
\vspace{30pt}
\vspace{30pt}
\vspace{30pt}
{\bf Mohd Ali \footnote{mohd.ali@students.iiserpune.ac.in}}, {\bf Vardarajan
Suneeta\footnote{suneeta@iiserpune.ac.in}}\\
\vspace{24pt} 
{\em  The Indian Institute of Science Education and Research (IISER),\\
Pune, India - 411008.}
\end{center}
\date{\today}
\bigskip
\begin{center}
{\bf Abstract}\\
\end{center}
In this paper, we show a local generalized second law (the generalized entropy is nondecreasing) in crossed product constructions for maximally extended static and Kerr black holes using modular theory. The new ingredient is the use of results from a recent paper discussing the entropy of the algebra of operators in subregions of arbitrary spacetimes. These results rely on an assumption which we show is true in our setting. However, we do assume as in that paper, that the gravitational constraints are implemented on each partial Cauchy slice. We employ a slight generalization of the construction, by including an observer degree of freedom even for wedge shaped regions with an asymptotic boundary. In the last part of the paper, we look at modular Hamiltonians of
deformed half-spaces in a class of static spacetimes, including the Schwarzschild spacetime. These are computed using path integrals, and we primarily compute them to investigate whether these non-local modular Hamiltonians can be made local by subtracting off pieces from the algebra and its commutant, as has been surmised in the literature. Along the way, the averaged null energy condition (ANEC) also follows in this class of spacetimes.

\newpage
\section{Introduction}
Bekenstein proposed the Generalized Second Law (GSL) \cite{JB}, \cite{JB2} for black hole spacetimes in order that the second law of thermodynamics be valid near black holes. This is the statement that the generalized entropy is nondecreasing, $\frac{dS_{gen}}{dv} \geq 0$, where $v$ is the null coordinate on the horizon. Here,
\begin{equation}\label{In1}
S_{gen} = <\frac{A}{4\hbar G}> + S_{QFT},
\end{equation}
where $A$ is the black hole horizon area at an arbitrary cut of the event horizon and $S_{QFT}$ is the entanglement entropy of the quantum fields in the black hole exterior. When Bekenstein originally proposed this, he had in mind the thermodynamic entropy of matter outside the horizon - it was Sorkin \cite{sorkin} who proposed that if the matter was a quantum field, $S_{QFT}$ should be the entanglement entropy of the exterior fields with the interior.

The GSL for Einstein gravity was proved by Wall \cite{AW} under an assumption of a renormalization scheme for the boost energy and the quantum field theory (QFT) entropy. As is also well-known, both the terms in (\ref{In1}) are individually ultraviolet (UV) divergent (the first term due to loop effects which renormalize $G$ and the second term, entanglement entropy, which is UV divergent), but there is a lot of evidence that the sum is UV finite \cite{SU}---\cite{gesteau}.
Recently, it was shown that a statistical quantity, the entropy of the algebra of observables in the AdS-Schwarzschild black hole exterior, was equal to the generalized entropy of the black hole at its bifurcation surface, modulo a state independent constant \cite{CPW}. The entropy of the algebra was defined using a crossed product construction by Witten \cite{W}. The algebra of QFT operators in the right exterior of the black hole is a Type  \RomanNumeralCaps {3}$_1$ von Neumann algebra \footnote{ The boundary version of this statement in the AdS/CFT correspondence was found by Leutheusser and Liu \cite{LL}, \cite{LL1} (see also \cite{lashkari}). They studied the holographic boundary operator algebra of the CFT dual to gravity in the asymptotically anti-de Sitter (AdS) black hole spacetime. They found a emergent Type  \RomanNumeralCaps {3}$_1$ von Neumann algebra for single trace operators in the large $N$ limit of the boundary CFT.} Later, in \cite{CLPW}, Chandrasekaran, Longo, Penington and Witten discussed how this construction can be generalized to asymptotically flat black holes. Considering quantum fields in the exterior of a Schwarzschild black hole, by including the ADM Hamiltonian in the set of operators and enlarging the Hilbert space to include the timeshift degree of freedom, one considers the crossed product of the original algebra with its modular automorphism group. This changes the algebra type to Type \RomanNumeralCaps {2}. The von Neumann algebras are classified based on their properties - for a review of von Neumann algebras and their classification, see \cite{EW1}. A Type \RomanNumeralCaps {2} von Neumann algebra (unlike a Type  \RomanNumeralCaps {3} algebra) has a notion of a (renormalized) trace, density matrix and thus, one can associate an entropy with the algebra, which is just the corresponding von Neumann entropy associated with the density matrix.The algebra entropy was shown to be the generalized entropy of the black hole at its bifurcation surface, modulo a state independent constant. These computations were generalized to black holes formed from collapse and Kerr black holes in \cite{JSG}. The entropy of the algebra has a monotonicity property under trace preserving inclusions, which can be used to prove a version of the Generalized Second Law (GSL) for asymptotically $AdS$ black holes showing that generalized entropy increases between early and late times when the separation between these times is large \cite{CPW}. This is not quite a \emph{local} GSL $\frac{dS_{gen}}{dv} \geq 0$ in crossed product constructions. A similar version of the GSL has been shown for an arbitrary diffeomorphism invariant theory of gravity in \cite{MAS}.

Our primary goal in this paper will be to show that a \emph{local} GSL is indeed true in crossed product constructions for maximally extended static black holes and maximally extended Kerr. We also discuss the asymptotically $AdS$ black holes in section V. To do this, we need to utilize constructions due to Jensen, Sorce and Speranza (JSS) in \cite{KJA}. They studied the algebra of operators associated to domains of dependence of arbitrary partial Cauchy slices in Einstein gravity plus matter. These algebras are Type \RomanNumeralCaps {3}. But it is possible to convert them to Type  \RomanNumeralCaps {2}, by doing a crossed product by the modular automorphism group. Their work is based on a conjecture on the modular Hamiltonian used in the construction. The conjecture is that the gravitational Hamiltonian generating the flow of a special vector field on the Cauchy slice, which is a local integral on the Cauchy slice, is the modular Hamiltonian for some state. As support for this conjecture, JSS argue that it should be possible to start with a non-local modular Hamiltonian and subtract appropriate terms and make it local. Then the converse of the Connes cocycle theorem states that this local integral is indeed the modular Hamiltonian for some state. With this conjecture, it is possible to associate an algebra entropy with the subregion, and show that this is the generalized entropy of the subregion modulo a constant. In the setting of our application of the JSS results, their conjecture turns out to be true. By considering a slight generalization of the JSS construction to include an observer even for wedge shaped regions with an asymptotic boundary, we have shown that we obtain a local GSL. We also give evidence for this conjecture for more general modular Hamiltonians in Appendix B - specifically seeing how a non-local modular Hamiltonian can be made local by subtracting off appropriate terms. For this, we need modular Hamiltonians in more general settings such as arbitrary wedge-shaped regions in curved spacetimes. These are hard to compute except when there is a high degree of symmetry. We compute modular Hamiltonians using the path integral method for perturbed Cauchy slices in a class of static spacetimes which includes the Schwarzschild spacetime, to linear order in the perturbation. This uses the methods of Faulkner, Leigh, Parrikar and Wang (FLPW) in \cite{TROH} and we use these modular Hamiltonians to discuss the conjecture of JSS.

Gravitational crossed product constructions have been explored in \cite{bahiru}. For crossed product constructions without gravity, and for a connection between the crossed product and extended phase space, see \cite{leigh}, \cite{leigh1}. Crossed product constructions for quantum field theories on subregions are discussed in \cite{jefferson}. Approximations to the crossed product for Type  \RomanNumeralCaps {1} algebras are explored in \cite{soni}. The use of the crossed product as a covariant regulator is discussed in \cite{speranza}.

The outline of the paper is as follows: In section 2, we define a half-sided modular inclusion, which is used crucially to obtain modular Hamiltonians in black hole spacetimes. In section 3, we discuss modular Hamiltonians in black hole spacetimes, both static and Kerr black holes. In section 4, we review the salient results of JSS on algebra entropy of subregions of spacetime which are domains of dependence of partial Cauchy slices. In section 5, we derive a local GSL using these results. In Appendix A, we discuss modular Hamiltonians of wedges in Minkowski spacetime as a warm-up example for the application of half-sided modular inclusions. In Appendix B, we compute (one-sided) modular Hamiltonians for various subregions of a general class of spacetimes and discuss whether they satisfy the conjecture of JSS. In the Discussion section, we provide a summary and discussion of our results.

\section{The half-sided modular inclusion (HSMI)}
In this section, we want to introduce the definition and properties of the half-sided modular inclusion (HSMI).
Let $A$ be a von Neumann algebra acting on the Hilbert space $\mathcal{H}$, with the cyclic and separating vector $\Omega \in \mathcal{H}$. The modular operator for $A$ is $\Delta_A$.\\
1) $\mathcal{H}_{smi}(A)^-$ is a von Neumann sub-algebra $B$ of $A$ with the properties:
\hspace*{10mm}  a) $\Omega$ is cyclic and separating for $B$.
\\
\hspace*{10mm}  b) $\Delta_A^{it} B \Delta_A^{-it} \subset B$ for $t\leq 0$.
\\
In this case $B$ is called the positive half-sided modular inclusion of $A$.\vspace{3mm}
\\
2) $\mathcal{H}_{smi}(A)^+$ is a von Neumann sub-algebra $B$ of $A$ with the properties:
\hspace*{10mm}  a) $\Omega$ is cyclic and separating for $B$.
\\
\hspace*{10mm}  b) $\Delta_A^{it} B \Delta_A^{-it} \subset B$ for $t\geq 0$.
\\
In this case $B$ is called the negative half-sided modular inclusion of $A$.
\\
Let $\Delta_B$ be the modular operator of $B$. There is a theorem ensuring the existence of the one-parameter continuous unitary $U(t)$ such that $U(1)$ maps $A$ and $B$ \cite{HJB}.
\\
\textbf{Theorem 1:} If $A$ and $B$ are von Neumann algebras such that $B$ is the half-sided modular inclusion of $A$, then there exists a one parameter continuous unitary $U(t)$ with $t \in \mathbb{R}$ with the following properties:
\\
\textbf{When inclusion is negative}
\\
\hspace*{10mm} a) $U(t)$ has a positive generator, i.e we can write
\begin{equation}\label{A1}
U(t)= \exp[iHt], \hspace{3mm} \text{with} \hspace{7mm} H\geq 0
\end{equation}
\hspace*{10mm} b) $U(t) \Omega=\Omega$   $\forall  t\in \mathcal{R}$
\\
\hspace*{10mm} c) $U(t) A U(-t) \subset A$  and $t\leq 0$
\\
\hspace*{10mm} d) $B= U(-1) A U(1)$
\\
\hspace*{10mm} e) $\Delta_A^{-it}\Delta_B^{it}=U(1-e^{-2\pi t})$.
\\
\hspace*{10mm} h)  $\Delta_A^{it} U(s) \Delta_A^{-it}= U(e^{(2\pi t)}s)$
\\
\textbf{When inclusion is positive}
\\
\hspace*{10mm} a) $U(t)$ has a positive generator, i.e, we can write
\begin{equation}\label{1A1}
U(t)= \exp[iHt], \hspace{3mm} \text{with} \hspace{7mm} H\geq 0
\end{equation}
\hspace*{10mm} b) $U(t) \Omega=\Omega$   $\forall  t\in \mathcal{R}$
\\
\hspace*{10mm} c) $U(t) A U(-t) \subset A$  and $t\geq 0$
\\
\hspace*{10mm} d)$B= U(1) A U(-1)$
\\
\hspace*{10mm} e) $\Delta_A^{-it}\Delta_B^{it}=U(e^{2\pi t}-1)$.
\\
\hspace*{10mm} h) $\Delta_A^{it} U(s) \Delta_A^{-it}= U(e^{(-2\pi t)}s)$
\vspace{5mm}
\\
The conditions a,b and c in (\ref{A1}) and (\ref{1A1}) define what is known as half-sided modular translation in the literature. As a warm up example, we have shown in the appendix \ref{app1} how one can obtain the modular operator for arbitrary wedges in Minkowski spacetime using modular inclusions. In particular, we obtain the modular operator of the wedge which is a translated version of the original wedge at the origin by constant amount. One can also obtain the modular operator of a null translated wedge where the null translation depends on the transverse coordinate. The latter case is a simpler case of a null translated wedge in a black hole spacetime which will be used while proving the GSL.
\section{Half-sided modular inclusions in black hole spacetimes}
\begin{figure}[h]
  \centering
  \includegraphics[width=0.90\textwidth]{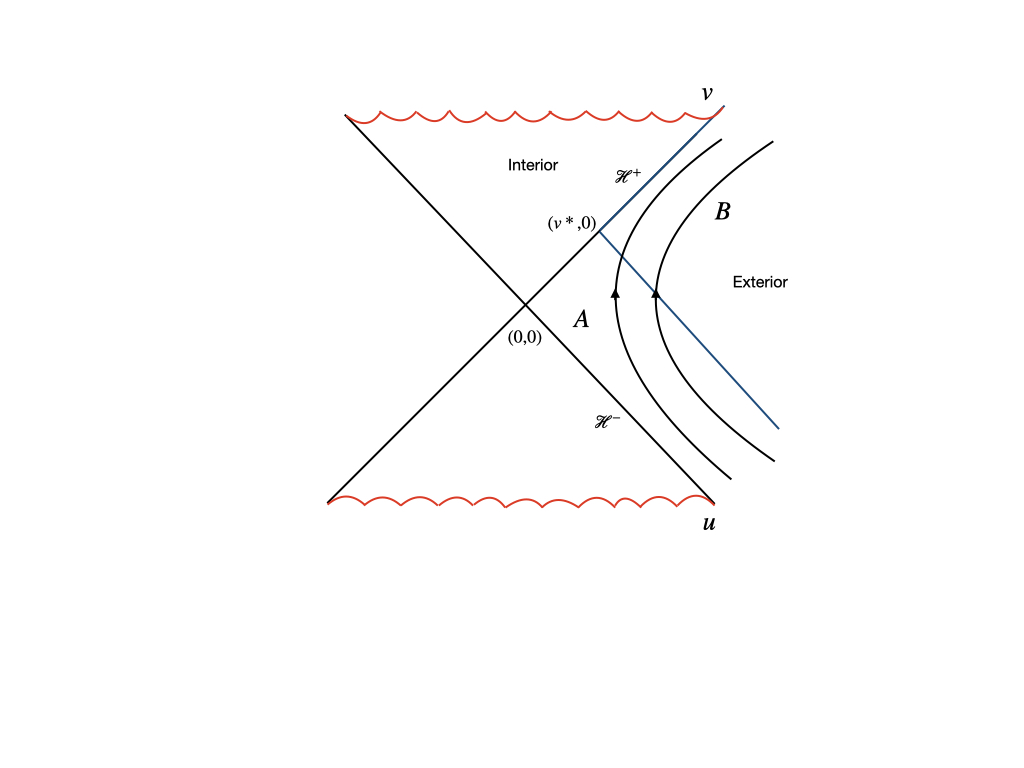}
  \caption{The figure depicts a black hole spacetime with Kruskal-like coordinates.  }
  \label{fig:BHW1}
\end{figure}

In this section, we want to obtain the relation between the modular Hamiltonian of two wedges in both static and Kerr black holes. For static black holes, the wedges are $A$ and $B$, as shown in Figure \ref{fig:BHW1}. We will restrict ourselves to black holes with a bifurcate Killing horizon and a smooth bifurcation surface. Both asymptotically flat and asymptotically $AdS$ black holes are considered.  We will also obtain such a relation for the two wedges on $\mathcal{H^+}$ for the Kerr black hole.
\\
\textbf{Static Black holes:}
\\
Let $\mathcal{M_A}$ and $\mathcal{M_B}$ be the von Neumann algebras associated with the wedge $A$ and $B$, respectively, as shown in Figure \ref{fig:BHW1}. Here, $v^{*}(y) > 0$ can be a function of the transverse coordinates $y$. Sewell's work in \cite{GLS} established that the modular Hamiltonian of the right exterior in the static black hole spacetime generates time translation with respect to the asymptotically timelike Killing vector in the right exterior (for e.g, the Schwarzschild time in a Schwarzschild black hole). The time translation Killing field behaves like a boost vector field on the event horizon of the black hole. Let $\xi^\mu$ be the Killing field associated with the time translation in this spacetime and let $T_{\mu \nu}$ be the stress tensor of all quantum fields present in this spacetime. Then, let us define
\begin{equation}\label{B1}
K_A=\frac{2\pi}{\kappa}\int_{\Sigma}T_{\mu \nu}\xi^{\mu}d\Sigma^{\nu}
\end{equation}
where $\Sigma$ represents a Cauchy surface in spacetime and $\kappa$, the surface gravity, is specific to the black hole. In this spacetime, there exists a Hartle-Hawking state $\Omega_{HH}$ that is a unique stationary state with respect to Killing time and is regular at the horizon \cite{BR}. This state is also KMS when restricted to the wedge $A$. The KMS condition for operators $a$ and $b$ in $\mathcal{M_A}$ is
\begin{equation}\label{B1A}
<\alpha_t (a)b > = <b \alpha_{t+ i\beta}(a)>
\end{equation}
Here, $\alpha_t $ is the automorphism of the algebra generated by the isometry of translations of the time (generated by the asymptotically timelike Killing vector). The modular operator for $(\mathcal{M_A},\Omega_{HH})$ for the right exterior is $\Delta_A = \exp[-K_A]$. Modular flow $\Delta_A^{it}=\exp[-iK_A t]$, corresponds to boost-like flow near the horizon and timelike flow inside $A$. $\alpha_t(a) = \exp[-iK_A t] a \exp[+ iK_A t]$, and we know from modular theory that this is an automorphism of the algebra. Now, since $\mathcal{M_B} \subset \mathcal{M_A}$, this implies $\Omega_{HH}$ is separating for $\mathcal{M_B}$. Furthermore, the fact that  $\Omega_{HH}$ is cyclic with respect to $\mathcal{M_A}$ and the spacetime possesses a global timelike Killing field implies $\Omega_{HH}$ is cyclic for $\mathcal{M_B}$ \cite{AS,ARM}.
In this section, we will choose for the Cauchy slice $\Sigma$ of the black hole right exterior, $\mathcal{H}^+ \cup \mathcal{I}^+$, the union of the future event horizon and future null infinity (both on the left and right for the maximally extended black hole).
\vspace{3mm}
\\
As seen in Figure \ref{fig:BHW1}, $\Delta_A^{it}$ has a local geometrical action on the operators in $\mathcal{M_B}$, causing them to move along integral curves of the boost Killing field. Since the flow is null on the horizon and timelike inside, the forward boost cannot take the local operator in $\mathcal{M_B}$ outside it. This implies: $\Delta_A^{it} \mathcal{M_B} \Delta_A^{-it} \subset \mathcal{M_B}$ for $t \leq 0$. Therefore, according to the definition of positive half-sided modular inclusion, $\mathcal{M_B}$ is the positive half-sided modular inclusion (HSMI) of $(\mathcal{M_A}, \Omega)$. Once the inclusion holds, Theorem 1 guarantees the existence of a unitary $U(t)$ such that
\begin{equation}\label{B2}
\Delta_A^{-it}\Delta_B^{it}=U(e^{2\pi t}-1)
\end{equation}
where $U(t)= \exp[i \mathcal{E}_{v*}t]$ where $\mathcal{E}_{v*}$ is a positive operator. From property (d) of a positive half-sided modular inclusion, it follows that $\mathcal{E}_{v*}$ is the generator of a null translation and can be written as
\begin{equation}\label{B2A}
\mathcal{E}_{v*}=  \int_{-\infty}^{\infty} \int d^2 \Omega~ dv~ v^{*}(y) T_{vv}
\end{equation}
Now following the same steps as in the appendix (\ref{app1}) for the Minkowski wedges, we can easily show that
\begin{equation}{\label{B3}}
K_B= K_A-2\pi \mathcal{E}_{v*}
\end{equation}
and the modular flow of the wedge $B$
\begin{equation}\label{B4}
  \Delta_B^{it}= e^{-i K_B t}.
\end{equation}
We know there is no global translation symmetry in this spacetime, but null translation is a symmetry on the horizon. As a result, we anticipate that the modular Hamiltonian of the wedge $B$ will be expressed in local form, at least on the horizon (by local form, we mean as a local integral over the three dimensional Cauchy slice, which in this case is a portion of the horizon). If one chooses any other partial Cauchy slice in the wedge $B$ other than the horizon ($ \mathcal{H}^+ \cup \mathcal{I}^+$ for asymptotically flat black holes), the modular Hamiltonian will be non-local. It is straightforward to verify that the one-sided modular Hamiltonian when computed on  $\mathcal{H}_{v*}^{+} \cup \mathcal{I}^+$ in asymptotically flat spacetime and $\mathcal{H}_{v*}^{+} $ in $AdS$, is identical to the form of a conjectured modular Hamiltonian in \cite{KJA}, where $\mathcal{H}_{v*}^{+} $ is the portion of the future horizon for $v > v*$.
This is because
\begin{equation}\label{B4A}
K^{\mathcal{H}_{v*}^{+}\cup \mathcal{I}^+}_B= 2\pi\int_{v*}^{\infty} dv~\int d^2 \Omega~ T_{vv}(v-v^{*}) + K({\mathcal{I}^+}).
\end{equation}
Here, $ K({\mathcal{I}^+})$ is the contribution to the modular operator from the partial Cauchy slice ${\mathcal{I}^+}$, which is common to $K_B$ and $K_A$. Note that the first term on the right in (\ref{B4A}) is exactly of the form of a local modular operator proposed in \cite{KJA} in a perturbative gravity expansion where $T_{ab}$ also contains the stress-energy of gravitons. In \cite{KJA}, it is conjectured that for the domain of dependence of a partial Cauchy slice $\Sigma_A$ in any spacetime, an expression such as
\begin{equation}
\int_{\Sigma_A} T_{ab} V^{a} d\Sigma^b
\label{B4AA}
\end{equation}
is proportional to the modular Hamiltonian for some state, if the vector field $V$ in the integral obeys certain properties: it acts like a boost close to the entangling surface separating the partial Cauchy slice from its complement, and has a certain prescribed form on the domain of dependence of the complement of this Cauchy slice. Further, the vector $V$ is such that $\nabla_a V_b |_S = \kappa n_{ab}$ where $S$ is the entangling surface and $n_{ab}$ is the binormal to the surface satisfying $n_{ab}n^{ab} = -2$. $\kappa$ is a constant. It can be checked that the vector field $V = \kappa(v-v^*) \frac{\partial}{\partial v}$ satisfies all the requirements of \cite{KJA} on the horizon --- further, it can be suitably extended both to the rest of the wedge (by choosing for example $\kappa((v-v^*) \frac{\partial}{\partial v} - u \frac{\partial}{\partial u })$) and to the domain of dependence of the complement to the Cauchy slice. $\kappa$ is exactly the surface gravity on the horizon. The condition $\nabla_a V_b |_S = \kappa n_{ab}$ is satisfied if we take $v^{*}$ to be a constant. When $v^{*} = f(y)$, then we need to modify the condition in \cite{KJA} to be
\begin{equation}\label{B4B}
n^{ab} \nabla_a V_b |_S = -2 \kappa.
\end{equation}
As it happens, all the derivations in \cite{KJA} go through with this modification - further, it is possible to change coordinates and satisfy the condition $\nabla_a V_b |_S = \kappa n_{ab}$ . Now, if we have $v^{*}(y)$ being a non-trivial function of the transverse coordinates $y$, then $\nabla_a V_b$ will not get any contributions from the terms proportional to the Christoffel symbols since those terms are proportional to the components of $V$ which vanishes at the entangling surface $S$. The terms of the form $\partial_a V_b$ will contain derivatives with respect to the transverse coordinates. However, those terms get projected out when multiplied by the binormal $n^{ab}$ in (\ref{B4B}). So, $V$ satisfies (\ref{B4B}).

\begin{figure}[h]
  \centering
  \includegraphics[width=0.90\textwidth]{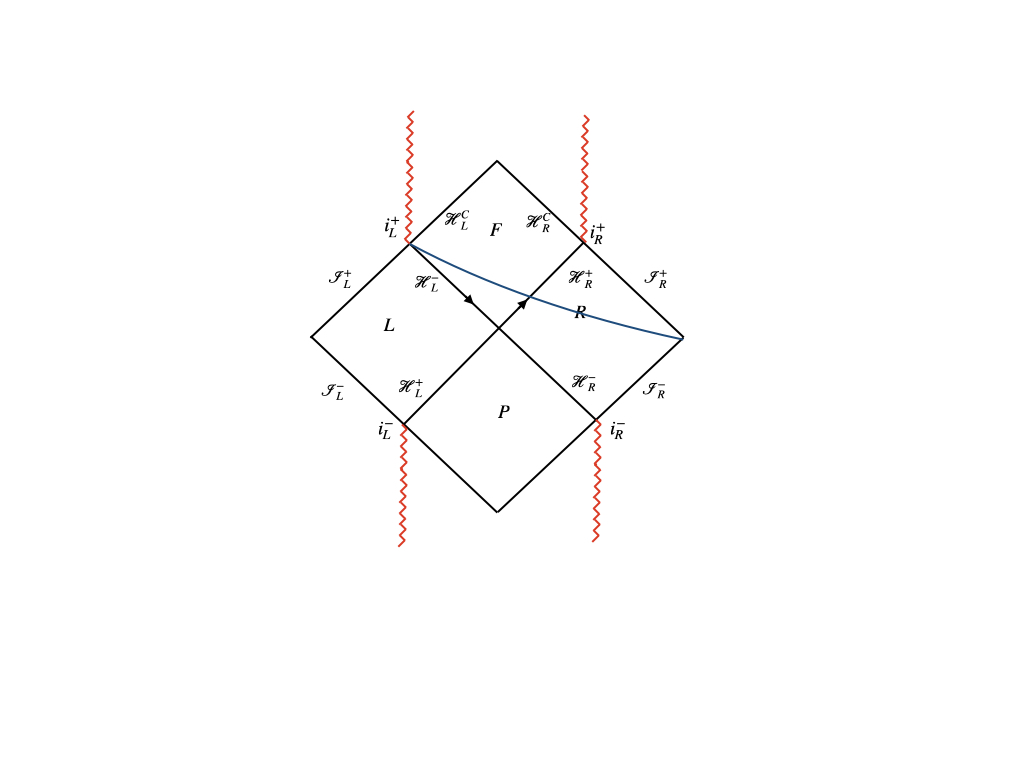}
  \caption{The figure depicts the Kerr black hole, where the red curvy line represents the singularity and the blue line represents the Cauchy surface for $R \cup F$.   }
  \label{fig:Kerr1}
\end{figure}

\begin{figure}[h]
  \centering
  \includegraphics[width=0.90\textwidth]{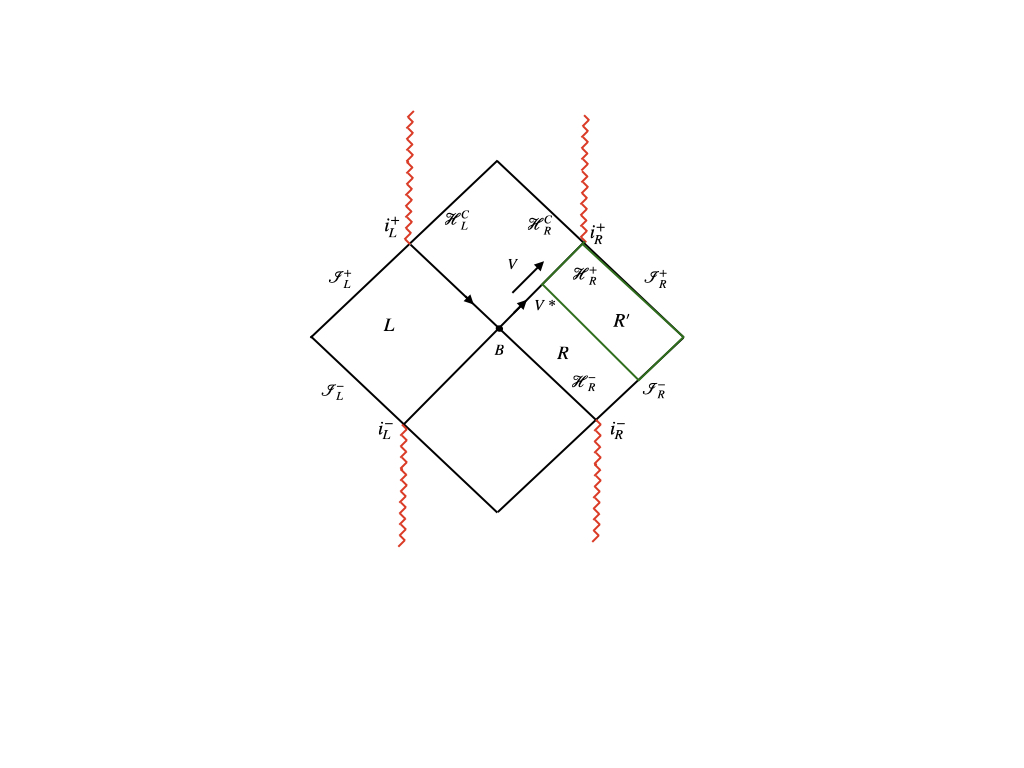}
  \caption{In this diagram, the green wedge represents bulk wedge $R'$, with the vertex at $V=V*$.   }
  \label{fig:Kerr2}
\end{figure}

Here, we know exactly that $K_B$ is the modular Hamiltonian of wedge $B$ for the \emph{same} state $\Omega_{HH}$ as $K_A$ and is obtained via the half-sided modular inclusion. Thus, this provides an example where the conjecture of JSS in \cite{KJA} is exactly true --- further, in their conjecture, the state for which this expression is the modular Hamiltonian is not known in general, whereas in this example, we know this.
\\
\\
\textbf{Kerr black hole:}
\\
The Kerr spacetime is stationary, and has a Killing horizon. The Killing field associated with the horizon is not timelike everywhere in the exterior of the black hole. As a result, there is no global KMS state on this spacetime. There is no Hartle-Hawking state for which we can repeat the procedure which we did for the Schwarzschild black hole. However, one can define a stationary state in the interior until the Cauchy horizon, and the exterior of the black hole \cite{WHU}. The Kerr spacetime is not globally hyperbolic when extended beyond the inner Cauchy horizon. Nevertheless, there is a well defined initial value problem for the exterior region $R$ and the region between the Cauchy horizon and the exterior horizon $F$. This has been discussed in great detail by Kudler-Flam, Leutheusser and Satishchandran (KLS) in the paper \cite{JSG}. Figure \ref{fig:Kerr1} depicts one of the Cauchy surfaces for these regions.
\\
 In this section, we will be primarily interested in the right exterior of the black hole. Let us consider linear fields in this spacetime which we generically denote $\Phi(x)$. First of all, it can be shown that any local field smeared with respect to some smooth function with compact support can be written in terms of smeared local fields on some Cauchy slice \cite{RMW}. We are interested in a Cauchy slice for the right exterior. It was shown by Wall \cite{AW} that one can take the Cauchy slice to be the union of ${\mathcal{H^+_R}}$ and $\mathcal{I}^+$.  We can construct the algebra of observables $\mathcal{A}_{\mathcal{H^+_R}}$ on ${\mathcal{H^+_R}}$ and $\mathcal{A}_{\mathcal{I^+_R}}$ on ${\mathcal{I}^+_R}$ and then the algebra of observables in the right exterior $\mathcal{A}_R \simeq \mathcal{A}_{\mathcal{H^+_R}} \otimes \mathcal{A}_{\mathcal{I^+_R}}$. This decomposition of bulk algebra in terms of boundary algebra has also been done by KLS in \cite{JSG} for the Cauchy slice which is a union of $\mathcal{H}^-$ and $\mathcal{I}^-$. KLS consider a Gaussian state in the black hole right exterior \footnote{These are states for which one-point functions vanish and $n$ point functions can be written as products of $2$ point functions.}. They assume that the state is Hadamard, stationary with respect to the horizon Killing field and has zero energy with respect to it. This state can be written as a state on the von Neumann algebra $\omega_{0}: \mathcal{A}_R \to C$. This state is denoted $\omega_0=\omega^{\mathcal{H}^-} \otimes \omega^{\mathcal{I}^-}$, where $\omega^{\mathcal{H}^-}$ and $\omega^{\mathcal{I}^-}$ are invariant with respect to the Killing field, and are Gaussian at $\mathcal{H}^-$ and $\mathcal{I}^-$ respectively. Then, they take the unique Gaussian state invariant under affine time translations on $\mathcal{H}^-$ and $\mathcal{I}^-$. They show this obeys the KMS condition (\ref{B1A}) on the horizon.
 This enables them to determine the form of the modular Hamiltonian on $\mathcal{H}^-$ and $\mathcal{I}^-$.
  \begin{equation}
 K_{-}=\frac{2\pi}{\kappa}\int_{\mathcal{H}^- \cup \hspace{1mm} \mathcal{I_R}^-}T_{\mu \nu}\xi^{\mu}d\Sigma^{\nu}
 \end{equation}
 where $\xi^\mu= t^\mu + \Omega_H \psi^\mu$. $t^\mu$ is the time translation Killing field and $\psi^\mu$ is the azimuthal Killing field in the Kerr spacetime.
 We can do the same at $\mathcal{H}^+$ and $\mathcal{I}^+$. Thus, we assume the existence of a quasifree Hadamard, stationary state with zero Killing energy and consider the unique state invariant under affine time translations on $\mathcal{H}_L^-$, $\mathcal{H}_R^+$ and $\mathcal{I}^+$. Such a state is automatically KMS on $\mathcal{H}_R^+$. The two-point function in this state is given on $\mathcal{H}_R^+$ with coordinates $(V, x^A)$ by
\begin{equation}\label{2point}
\omega_{0}(\Pi(x_1) \Pi(x_2)) = -\frac{2}{\pi}  \frac{\delta_{S^2}(x_{1}^{A}, x_{2}^{A})}{(V_1 - V_2) -i 0^{+}}.
\end{equation}
Here, $\Pi(x) = \partial_V \Phi$ are operators supported on $\mathcal{H}_R^+$.
Considering the geometric flow of the Killing time translation $(V, x^{A}) \to (e^{\kappa t}V, x^{A})$ on $\mathcal{H}^+$ --- this generates an automorphism of the algebra, $\alpha_t$. With respect to this flow, it can be checked that the state $\omega_{0}$ is KMS \cite{JSG}. A similar observation can be made at $\mathcal{I_{R}^{+}}$. Added to the assumption that this state has zero Killing energy on the horizon, this implies that this flow is modular flow and the modular Hamiltonian is given on $\mathcal{H}_L^- \cup \mathcal{H}_R^+ \cup \hspace{1mm} \mathcal{I_{R}^{+}}$ by
\begin{equation}\label{3point}
K=\frac{2\pi}{\kappa}\int_{\mathcal{H}_L^- \cup \mathcal{H}_R^+ \cup \hspace{1mm} \mathcal{I_{R}^{+}}} T_{\mu \nu}\xi^{\mu}d\Sigma^{\nu}
\end{equation}
 We note that the modular Hamiltonian on some other Cauchy slice may not have this nice, local form and away from this slice, modular flow may not be a geometric flow. We can formally split the above equation as an integral on $\mathcal{H}_R^+ \cup \mathcal{I}^+$ and $\mathcal{H}_L^-$ and call it $K_R$ and $-K_L$,
 \begin{equation}\label{4point}
 K= K_R -K_L
 \end{equation}
This split is just formal because the entanglement across the bifurcation surface is infinite. This give rise to a type III von Neumann algebra $\mathcal{U}(\mathcal{H}_R^+,\omega_0)$. Further, using the fact that for linear fields, one can always write any observable in the bulk region in terms of an observable on the Cauchy slice pertaining to that bulk, the algebra of the right exterior can be written as \footnote{ We assume that we can specify the initial data independently on $\mathcal{H}_R^+$ and $\mathcal{I}_R^+$.},
 \begin{equation}\label{B5}
 \mathcal{U}(R,\omega_0) \simeq \mathcal{U}(\mathcal{H}_R^+,\omega_0) \otimes  \mathcal{U}(\mathcal{I}_R^+,\omega_0)
 \end{equation}
  Further, the algebra on the $\mathcal{H}_L^-$, $\mathcal{U}(\mathcal{H}_L^-,\omega_0)$ is the commutant of $\mathcal{U}(\mathcal{H}_R^+,\omega_0)$. More details can be found in the paper of KLS \cite{JSG}.
  \\
 In the null coordinates on $\mathcal{H}^+$, the modular Hamiltonian (\ref{3point}) generates dilatation on the horizon, see Figure \ref{fig:Kerr1}. Consider another wedge  $R'$ that has one of its future null boundaries overlapping with the part of $\mathcal{H}_R^+$, as shown in Figure \ref{fig:Kerr2}.
\\
Let $V$ be an affine time on $\mathcal{H}^+$ which goes from $0$ to $\infty$ as we move from the bifurcation surface $B$ to $i^+_R$, as shown in the Figure \ref{fig:Kerr2}. Let the vertex of the new wedge be at $V*$ and $\mathcal{H}^+(V*)= \mathcal{H}^+ \cap (V \geq V*)$. Following the above discussion, the algebra of the wedge can be represented by the boundary algebra  $ \mathcal{U}(R',\omega_0) \simeq \mathcal{V}(\mathcal{H}^+(V*),\omega_0) \otimes  \mathcal{U}(\mathcal{I}_R^+,\omega_0)$. The state $\omega_0$ here is a restriction of $\omega_0$ to $R'$. The state $\Omega$ is clearly separating and cyclic \cite{AS} for $\mathcal{V}(\mathcal{H}^+(V*),\omega_0) $ by construction. Since modular flow produces dilatation on the horizon, $ \mathcal{U}(R',\omega_0)$ is the HSMI of the algebra $\mathcal{U}(R,\omega_0)$. Following the same steps as for the Schwarzschild and Rindler spacetime, the modular operator of $R'$ is,
\begin{equation}\label{B6}
K(V*)= K - 2\pi \mathcal{E}_{V*}
\end{equation}
where $K(V*)$ and $K$ the modular Hamiltonian for the wedge $R'$ and $R$ and $\mathcal{E}_{V*}$ is the generator of the null translation connecting them. This modular operator (\ref{B6}) is the modular operator for the state $\omega_0$ and it has a local geometric action on the boundary algebra, since the null translation is a symmetry on the horizon. It will not have a local geometrical action in the exterior $R'$ away form the horizon.
\section{Review: Type II construction for gravitational subregion}\label{S}
In this section, we would like to summarize the construction of Jensen, Sorce and Speranza (JSS) in \cite{KJA} which we will use to prove the GSL in the next section. This construction is for Einstein gravity coupled to quantum fields. The construction generalizes recent work which studies QFT in a static black hole background in \cite{W,CPW,CLPW}, where the entropy of an algebra is discussed. This algebra is that of fields in the black hole exterior. In \cite{W,CPW,CLPW}, this algebra is enlarged using the crossed product construction in von Neumann algebra. The operator that is added in the enlarged algebra is the ADM mass which now acts on an enlarged Hilbert space which includes square integrable functions of a new degree of freedom, the timeshift. This crossed product (by the modular automorphism group) changes the algebra of fields in the right exterior from a Type III$_1$ von Neumann algebra to a Type II algebra. Consequently, one has a renormalized trace on the algebra which can be used to obtain a von Neumann entropy in any state for the algebra. \cite{W,CPW,CLPW} showed that this entropy is the generalized entropy of the black hole at the bifurcation surface modulo a constant. In this construction, the isometry group of the static black hole is implemented as a set of constraints - the time translation generator then equals a boundary term --- the difference of the left and right ADM Hamiltonians.

In \cite{KJA}, JSS have generalized the crossed product construction of \cite{W,CLPW,CPW} for arbitrary subregions to obtain the entropy of the algebra of domains of dependence of partial Cauchy slices. This is for subregions in theories of Einstein gravity coupled to matter in the Newton's constant $G_N \rightarrow 0$ limit. The construction depends on a specific vector field and relies on the existence of a conjectured state whose modular flow is local and geometrical on some Cauchy slice.
\\
\textbf{JSS construction:}
\\
 \begin{figure}[h]
  \centering
  \includegraphics[width=0.90\textwidth]{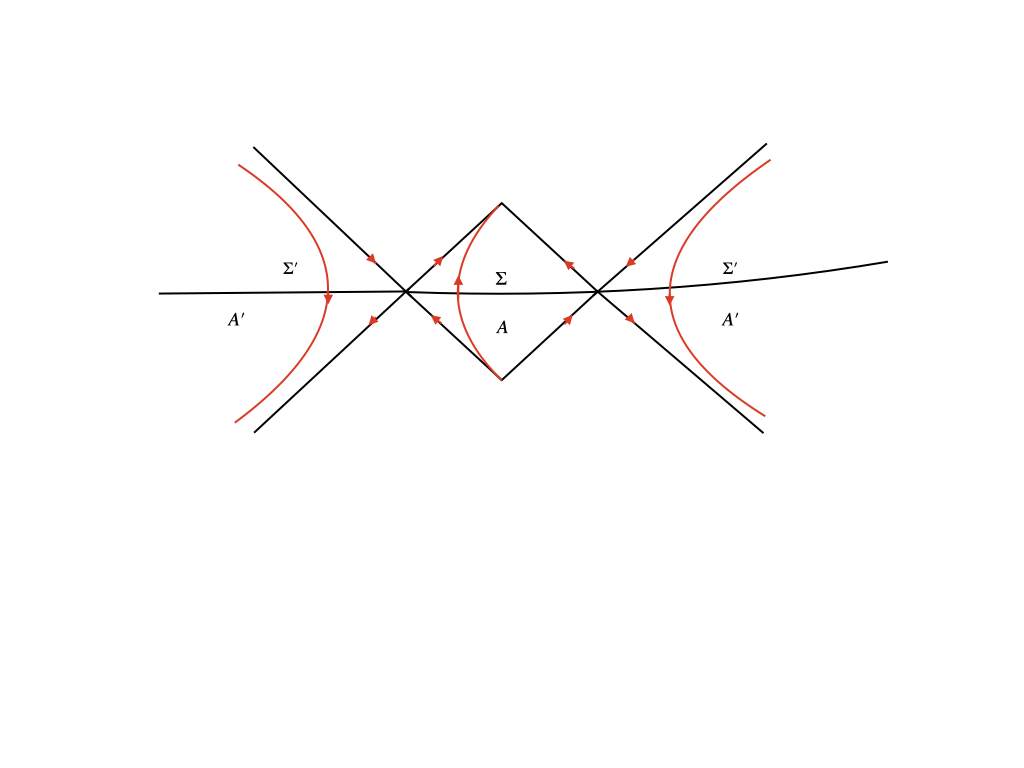}
  \caption{In this diagram A is the subregion of interest and A' is its complement. $\Sigma$ and $\Sigma$' are partial Cauchy slice s associated with A and A'. Red lines in the diagram represent the vector field $\xi^a$ and its direction. }
  \label{fig:SS1}
\end{figure}
Let $A$ be a subregion and  $A'$ be the causal complement as shown in the figure \ref{fig:SS1}.  Since the observables must be diffeomorphism invariant, they must commute with the constraint associated with these diffeomorphisms. In particular, JSS consider a class of subregion-preserving diffeomorphisms that act both on $A$ and $A'$, with the following properties:
\\
$1)$ They generate boosts around the entangling surface.
\\
$2)$ The vector field  $\xi^a$, which generates this diffeomorphism should be future directed in $A$ and past directed in $A'$, and should be tangent to the null boundaries of the subregions.
\\
$3)$ $\xi^a$ must vanish at the entangling surface $\partial \Sigma$ and have constant surface gravity $\kappa$ on $\partial \Sigma$, given by
\begin{equation}\label{S2}
\nabla_a \xi_b \overset{\partial \Sigma}{=} \kappa n_{ab}
\end{equation}
where $n_{ab}$ is the binormal to $\partial \Sigma$.
\\
The gravitational algebra is obtained by imposing the diffeomorphism as a constraint on the algebra of observables order by order in the full nonlinear theory of gravity. However, as shown in \cite{CLPW}, directly imposing constraints on $\mathcal{A}_{QFT}$ and $\mathcal{A}_{QFT}'$ trivializes the algebra. Instead, one must introduce an observer in the subregion and extend the algebra by adding Hamiltonian $H_{obs}= \hat{q}$ of the observer in the algebra of observables.  Also, one must extend the Hilbert space by tensoring the Hilbert space of the QFT with the observer Hilbert space $\mathcal{H}_{obs}=L^2 (\mathcal{R})$. When the subregion does not contain any asymptotic boundary \footnote{In all the cases considered in our paper, the subregion will contain an asymptotic boundary.}, we need an observer to define the location of subregion. We must also add an observer in $A'$, but since $A'$ contains an asymptotic boundary, the role of the observer is played by the ADM Hamiltonian $H_{ADM}$ \footnote{This needs to be generalized to include an observer even for subregions with an asymptotic boundary, as we discuss in the next section.}.  Now, the full algebra is $(\mathcal{A}_{QFT}V \mathcal{A}_{QFT}') \otimes \mathcal{A}_{obs}\otimes \mathcal{A}_{ADM}$, and it acts on the Hilbert space $\mathcal{H}=\mathcal{H}_{QFT }\otimes \mathcal{H}_{obs}\otimes \mathcal{H}_{ADM}$. It can be shown that the gravitational constraint associated with $\xi^a$ is given by
\begin{equation}\label{S3}
C[{\xi}]= H^g_\xi + H_{obs}+H_{ADM}.
\end{equation}
Here, $H^g_\xi$ is the operator generating the flow $\xi^a$ on the quantum field algebra $\mathcal{A}_{QFT}$ and $\mathcal{A}_{QFT}'$ instantaneously on the Cauchy slice $\Sigma_c =\Sigma \cup \Sigma'$ as shown in the figure \ref{fig:SS1}. One can write it as a local integral of the matter and graviton stress tensors (i.e., as an integral on the Cauchy surface), which at leading order has the form (linear order constraint having already been implemented):
\begin{equation}\label{S4}
H^g_\xi= \int_{\Sigma_c} T_{\mu \nu}\xi^\nu d\Sigma^\mu
\end{equation}
Further, JSS have shown that implementing the constraint at the level of the subregion algebra gives the von Neumann algebra $\mathcal{A}^C$, which is the crossed product of $\mathcal{A}_{QFT}$ by the flow generated by $H^g_\xi$.
\begin{equation}\label{S4A}
\mathcal{A}^C=\{ e^{iH^g_\xi \hat{p}} a  e^{-iH^g_\xi \hat{p} }, e^{i\hat{q}t}| a\in A_{QFT}, t\in\mathcal{R}\}''
\end{equation}
where $\hat{p}$ is the canonical conjugate to the observer Hamiltonian and $S''$ denotes the smallest von Neumann algebra containing the set $S$. \textbf{Further, JSS assume that $H^g_\xi$ is the modular Hamiltonian for some state on the algebra $\mathcal{A}_{QFT}$}. With this assumption, the algebra in (\ref{S4A}) becomes type II, for more details, see \cite{KJA}. The assumption that $H^g_\xi$ is a modular Hamiltonian is a key assumption for obtaining the type II algebra --- there are cases for which this assumption is \textbf{exactly true,} and these situations are the focus of our attention in the previous and next sections. Once the type II algebra is obtained, it is straightforward to define a renormalized trace and to associate entropy with the algebra as done in \cite{CLPW,CPW,W},
\begin{equation}\label{S5}
S(\rho_{\hat{\phi}})= -S_{rel}(\Phi||\Psi)- \beta \Big< H_{obs}\big>_f+ S^f_{obs}+\log{\beta}
\end{equation}
where $\beta$ is the inverse temperature associated with the KMS state $\Psi$ for which $H^g_\xi$ is the modular operator and  $\hat{\Phi}= \Phi \otimes f$ is the state in the crossed product construction of the type II algebra, see \cite{KJA}. Here, we have assumed that the state  $\hat{\Phi}$ is semiclassical. This is defined as a eigenstate of the conjugate momentum operator $\hat{p}$ peaked around zero momentum or, equivalently, a state with a slowly varying position wavefunction $f(\hat{q})$. This assumption is crucial because, in crossed product algebras, operators are dressed with respect to $e^{iH_\xi^g \hat{p}}$. When acting on states of the form $|\Phi\rangle \otimes |f\rangle$, this dressing induces entanglement between the observer and quantum field degrees of freedom whenever the modular energy is non-zero. For the state with non-zero modular energy, the semiclassical assumption ensures this entanglement remains small, and simplifies the entropy of the algebra $S(\rho_\Phi)$, enabling the derivation of equation (\ref{S5}). Further it can be shown that  Araki's Type III relative entropy of the state $\Phi$ with respect to state $\Psi$ is \cite{araki}
\begin{equation}\label{S6}
S_{rel}(\Phi||\Psi)= \beta \Big< H^\Sigma_\xi \Big>_{\phi}- S_\phi^{QFT}-\beta \Big< H^\Sigma_\xi \Big>_{\psi}+S_{\Psi}^{QFT}
\end{equation}
where $H^\Sigma_\xi= \int_\Sigma T_{\mu \nu}\xi^\nu d\Sigma^\mu $ is the one-sided modular Hamiltonian. Finally, JSS show that the algebra entropy is the generalized entropy modulo a  state independent constant. JSS assume that the gravitational constraint $C[\xi]=0$ holds locally on the partial Cauchy slice for the subregion. This allows them to obtain an integrated first law of local subregions,
\begin{equation}\label{S6A}
H^\Sigma_\xi+H_{obs}=-\frac{1}{16\pi G_N}\int_{\partial\Sigma} n_{ab}\nabla^a \xi^b= -\frac{\kappa}{2\pi} \frac{A_{\partial \Sigma}}{4G_N}
\end{equation}
  where $A_{\partial \Sigma}$ is the area of the entangling surface \footnote{This is exactly the place where $\nabla_a \xi_b \overset{\partial \Sigma}{=} \kappa n_{ab}$ is used, but notice that we will get same answer even if we assume the weaker condition $n^{ab}\nabla_a \xi_b \overset{\partial \Sigma}{=} -2\kappa$ . By changing coordinates, we can also get exactly the JSS condition $\nabla_a \xi_b \overset{\partial \Sigma}{=} \kappa n_{ab}$}. Further, in quantum theory, the constraint should be implemented as an operator equation, which will give
  \begin{equation}\label{S7}
 \Big< H^\Sigma_\xi\Big>_{\Phi} +\Big<H_{obs}\Big>_{f}=-\frac{\kappa}{2\pi} \Big<\frac{A_{\partial \Sigma}}{4G_N}\Big>_{\hat{\Phi}}.
  \end{equation}
 Using (\ref{S6A}) and (\ref{S7}), it can be shown that
  \begin{equation}\label{S8}
  S(\rho_{\hat{\phi}})=\Big<\frac{A_{\partial \Sigma}}{4G_N}\Big>_{\hat{\Phi}} + S_{\Phi}^{QFT}+S_f^{obs}+c
  \end{equation}
  where $c$ is a state independent constant. The above equation is the relation between the entropy of the algebra and generalized entropy for the subregion.
  \begin{equation}\label{S9}
  c=\log{\beta}-\beta \Big< H^\Sigma_\xi \Big>_{\psi}+S_{\Psi}^{QFT}
  \end{equation}
\section{The Generalized Second law (GSL)}
In this section, we will show that a local generalized second law holds true in the crossed product constructions for static and Kerr black holes with bifurcate Killing horizons and in Rindler spacetime. While we will do the computation for asymptotically flat black holes, it can be done for asymptotically $AdS$ black holes as well.  The idea is to use the construction of JSS \cite{KJA}, but with a modification. As mentioned in the previous section, their construction relies on an assumption that their proposed local Hamiltonian is the modular Hamiltonian of some state. In our computation of modular Hamiltonians, we have shown that the modular Hamiltonian obtained using HSMI of the algebra is for the Hartle Hawking state of static black holes and Minkowski vacuum for Rindler (and for the particular Gaussian state we assumed in the Kerr case) - furthermore, it precisely matches with the form of the JSS local Hamiltonian (\ref{B4AA}), when the Cauchy slice contains the future horizon. This is true for both wedges $A$ and $B$ in Figure \ref{fig:BHW1}. For wedge $B$, the modular Hamiltonian is (\ref{B4A}) which is of the form (\ref{B4AA}) with vector field $V = \kappa(v-v^{*}(y)) \frac{\partial}{\partial v}$ on the horizon. This satisfies nearly all of the conditions of JSS (when suitably extended to the rest of the wedge as described in previous sections and in the complement Cauchy slice) and the modified condition (\ref{B4B}) which, as we have already mentioned in previous sections, is all we need for the JSS construction to go through. Thus, we have the modular Hamiltonians of both wedge $A$ and $B$ corresponding to the \emph{same} state. They both have the JSS form. So the assumption of JSS for a local modular Hamiltonian is explicitly realized in this case. Our aim is to use the JSS construction along with the positivity of relative entropy to show $S_{gen}(\infty)\geq S_{gen}(v*)$ for any $v^{*}(y) \geq 0$. Then, using monotonicity of relative entropy, we can establish that  $S_{gen}(v**)\geq S_{gen}(v*)$ for $v^{**}(y) \geq v^{*}(y)$. This is a local GSL.
\begin{figure}[h]
  \centering
  \includegraphics[width=0.90\textwidth]{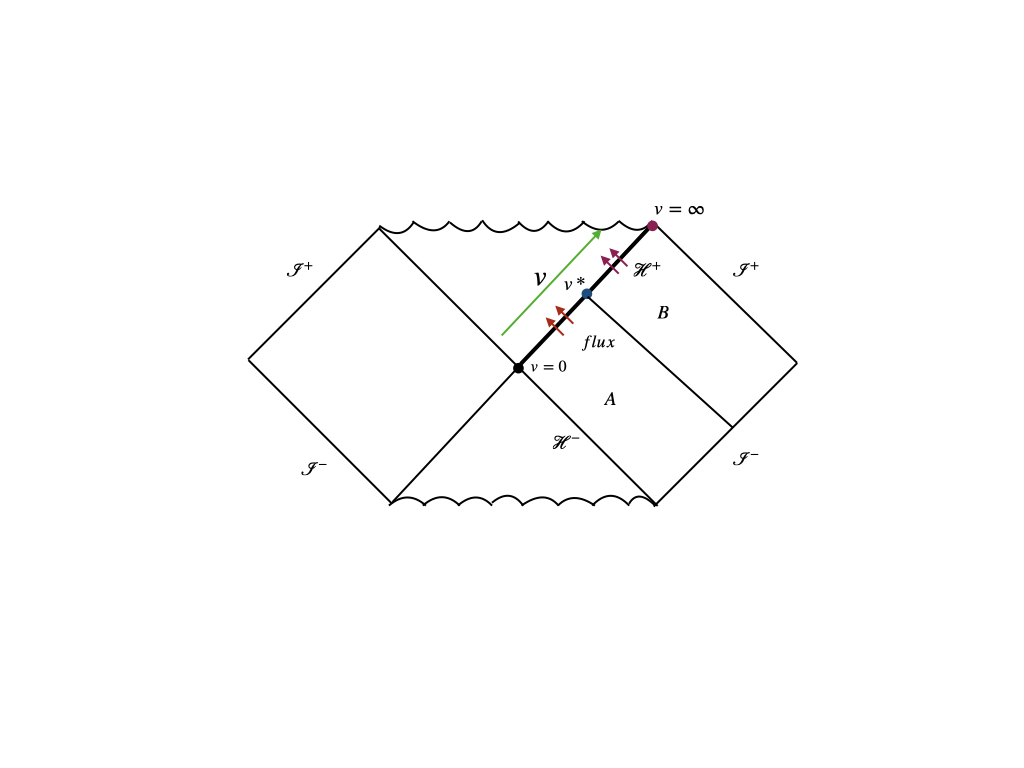}
  \caption{The figure represents accretion of quantum matter into the black hole.  }
  \label{fig:BHW2}
\end{figure}
\\
Before we get into the computation, we employ the JSS construction with a generalization which we describe in the next sub-section.
\subsection{Adding an observer to the calculation:}
We now wish to discuss a slight generalization of the JSS construction where we add an observer to the calculation. We will use the JSS construction to get relative entropy in two different wedges. For wedges with asymptotic regions, in the JSS construction, the ADM Hamiltonian associated with the chosen vector field plays the role of an observer in section 4. However, since this is a boundary term on a codimension 2 surface at infinity, it only depends on the asymptotic form of the vector field which is the same for the two wedges we are considering (or indeed for any wedge-shaped regions). To make the construction and dressing of operators specific to the wedges we are considering, we add an observer with Hamiltonian $H_{obs} = q \geq 0$ \emph{in addition to} $H_{ADM}$. The commutant has an observer with Hamiltonian $H'_{obs} = q' \geq 0$. This has also been discussed in \cite{JSG}. To begin with, we then have the Hilbert space $\mathcal{H}=\mathcal{H}_{QFT }\otimes \mathcal{H}_{obs}\otimes \mathcal{H'}_{obs} \otimes \mathcal{H}_{ADM} \otimes \mathcal{H}_{ADM'}$. Then, the constraint is
\begin{equation}\label{Discussion1}
C[{\xi}]= H^g_\xi + q - q'- H_{ADM} + H_{ADM'}.
\end{equation}
Properly implementing the constraint at the level of the Hilbert space \cite{CLPW} leads to the Hilbert space $\mathcal{H}=\mathcal{H}_{QFT }\otimes \mathcal{H}_{obs} \otimes \mathcal{H}_{ADM} \otimes \mathcal{H}_{ADM'}$.
We now dress operators with respect to the \textbf{observer} to make the dressing specific to the wedge. This produces the crossed product algebra
$\mathcal{A}^C$, which is the crossed product of $\mathcal{A}_{QFT}$ by the flow generated by $H^g_\xi$.
\begin{equation}\label{Discussion2}
\mathcal{A}^C=\{ e^{iH^g_\xi \hat{p}} a  e^{-iH^g_\xi \hat{p} }, e^{i\hat{q}t}, H_{ADM} | a\in A_{QFT}, t\in\mathcal{R}\}''
\end{equation}
where $\hat{p}$ is the canonical conjugate to the observer Hamiltonian and $S''$ denotes the smallest von Neumann algebra containing the set $S$. The inclusion of both the observer and the ADM Hamiltonian still leads to a Type II$_{\infty}$ von Neumann algebra as discussed in \cite{JSG}.
 Considering the states $\hat{\Phi}=  \Phi \otimes g \otimes f$ and $\hat{\Omega}= \Omega \otimes g \otimes f$ (the second factor relates to the observer degree of freedom, and the third factor to the $H_{ADM}$ degree of freedom), we can now discuss relative entropies of the two wedges. We note that implementing the constraint on each partial Cauchy slice now yields
\begin{equation}\label{Discussion3}
\Big<K_A(v*)\Big>_{\Phi} + \beta \Big< q \Big>_g- \beta\Big<H_{ADM}\Big>_f = - \Big<\frac{A}{4G_N}\Big >_{\hat{\Phi}}
\end{equation}
First, we will discuss the static black hole case. We have a static black hole with Killing horizon which is perturbed away from stationarity by  quantum fields (including gravitons) in the spacetime. We will assume that the black hole settles down to a stationary state at late times. This is plausible because all the flux of matter would either have crossed the horizon or would have escaped to future null infinity. In the case of $AdS$, all matter will eventually cross the horizon. So, at late times, the state must be indistinguishable from the vacuum $\Omega_{HH}$.  Let us now apply the modified JSS construction to the wedge $B$ for the black hole. This will yield a type II von Neumann algebra of fields in the wedge $B$. Once we have obtained the type II algebra, the modular Hamiltonian can be factorized and we can write a one-sided modular Hamiltonian. Further, using the definition of Araki's relative entropy, we can write $S(\hat{\Phi}||\hat{\Omega})$ in terms of the one-sided modular operator in the type II algebra, with the states $\hat{\Phi}=  \Phi \otimes g \otimes f$ and $\hat{\Omega}= \Omega \otimes g \otimes f$ in the type II algebra corresponding to some quantum field state $\Phi$ and cyclic, separating state $\Omega$ as defined in the previous section. Using (5.13) and (5.14) in the paper \cite{KJA} and the fact that we are working with a semiclassical state \footnote{One needs to use the fact that in a semiclassical state, $g(q)$ is slowly varying or equally its Fourier transform is peaked around zero momentum.}, it can be shown that $S(\hat{\Phi}||\hat{\Omega})$= $S(\Phi||\Omega)$. This explicitly shows the well-known fact that Araki's relative entropy is well-defined and finite even in a type III algebra. It allows us to write the relative entropy in terms of the one-sided modular Hamiltonian in the type III algebra (which is well-defined as a Hermitian form \cite{speranza}).  The partial Cauchy slice on which we want to write the modular Hamiltonian is $\mathcal{H}^+(v*) \cup \mathcal{I}^+$, where  $\mathcal{H}^+(v*)$ represents $v\geq v* $ part of the horizon. The modular Hamiltonian on this partial Cauchy slice can be written as $K_A(v*)= K_{\mathcal{H}^+}(v*) +  K_{\mathcal{I}^+}$. Here, $ K_{\mathcal{I}^+}$ is the modular Hamiltonian at $\mathcal{I}^+$ and
\begin{equation}\label{C1}
K_{\mathcal{H}^+}(v_*)= 2\pi \int_{v_*}^{\infty} dv \int_{\mathcal{H}^{+}} d^{D-2}x \sqrt{h} (v-v_*) T_{vv}
\end{equation}
which can easily be obtained using the equation (\ref{B3}) and the fact that $\mathcal{E}_{v*}$ is a generator of null translation and is local on the horizon.  Note that $K_{\mathcal{I}^+}$ is independent of $v*$. This implies that $K_{\mathcal{I}^+}$ will not contribute to the difference in the modular Hamiltonian between two cuts. The state of the system restricted to the wedge $B$ can be obtained by specifying the density matrix at $\mathcal{H}^+(v*)$ and $\mathcal{I}^+$,i.e  $\Psi(\mathcal{H}^+(v*) \cup \mathcal{I}^+)= \rho_{\mathcal{H}^+(v*)}\otimes \sigma_{\mathcal{I}^+}$ \cite{AW}.  Let $\Phi$ be the state of the quantum field, which is indistinguishable from $\Omega_{HH} $ at late times. Now as shown by JSS in \cite{KJA},
\begin{equation}\label{C2}
S_{rel}(\Phi||\Omega_{HH})=   \Big<K_A(v*)\Big>_{\Phi} -S_{\Phi}^{QFT} -\Big<K_A(v*)\Big>_{\Omega_{HH}}+S_{\Omega_{HH}}^{QFT}
\end{equation}
 where $S_{rel}(\Phi||\Psi)$ is Araki's relative entropy of the state $\Phi$ and $\Omega_{HH}$, $S_{\Phi}^{QFT}$ and $S_{\Omega_{HH}}^{QFT}$ are the entropy of the QFT in the state $\Phi$ and $\Omega_{HH}$ respectively. While each term may not be finite, but since relative entropy in type II and type III algebras are equal, all divergent terms must come in pairs in such a way that the final answer is finite. Further, the one-sided modular Hamiltonian is well-defined as a sesquilinear form on a dense set of states \cite{speranza}.

Imposing constraints as an operator equation and for some state $\Phi$,
\begin{equation}\label{C3}
\Big<K_A(v*)\Big>_{\Phi} + \beta \Big< q \Big>_g- \beta\Big<H_{ADM}\Big>_f = - \Big<\frac{A}{4G_N}\Big >_{\hat{\Phi}}
\end{equation}
where $\beta$ is inverse temperature associated with the KMS state $\Omega_{HH}$ and $\hat{\Phi}=  \Phi \otimes g \otimes f$ is the state in the crossed product construction of the type II algebra, see \cite{KJA} .  Similarly, $\hat{\Omega}= \Omega \otimes g \otimes f$. The $f$ and $g$ in the type II state are square integrable wavefunctions and $\beta\Big<H_{ADM}\Big>_f$ is the expectation value of the ADM Hamiltonian in $f$ while $\beta \Big< q \Big>_g$ is the expectation value of the observer Hamiltonian in state $g$. Notice that $f$ and $g$ are independent of the state $\Phi$ and therefore if we consider the difference of the area operator in two different quantum field states, than it will be independent of both $f$ and $g$. Finally, we see that the relative entropy of the states  $\hat{\Phi}$ and $\hat{\Omega}$ is the same as the type III entropy, and the terms dependent on the observer and ADM degrees of freedom \emph{cancel out in a single wedge}\footnote{We could think of the addition of the observer as a regularization tool since the observer degrees of freedom do not play a role in the relative entropy. The crossed product with respect to the observer as a regularization tool has also been discussed for QFTs in \cite{gomez}}. We will be doing this computation for two wedges that we define below, and inclusion of the observer degree of freedom explicitly defines the wedges.
Using (\ref{C3}) and (\ref{C2}), we get
\begin{equation}\label{C4}
S_{rel}(\Phi||\Omega_{HH})= \Big<\frac{A}{4G_N}\Big >_{\Omega_{HH}}-\Big<\frac{A}{4G_N}\Big >_\Phi + S_{\Omega_{HH}}^{QFT}-S_{\Phi}^{QFT}
\end{equation}
Since $\Omega_{HH}$ is a stationary state, the area at any cut is the same as at $v* \rightarrow \infty$. Using this fact, we can write the equation as
\begin{equation}\label{C5}
S_{rel}(\Phi||\Omega_{HH})(v*)= \Big<\frac{A}{4G_N}\Big >(\infty)-\Big<\frac{A}{4G_N}\Big >_{\Phi} (v*) + S_{\Omega_{HH}}^{QFT}-S_{\Phi}^{QFT}
\end{equation}
We put $v*$ in the equation above to emphasize that it is for the wedge at $v*$. Further, the above equation can also be written as:
\begin{equation}\label{C6}
S_{rel}(\Phi||\Omega_{HH})(v*)= S_{gen}(\infty)-S_{gen}(v*).
\end{equation}
The positivity of relative entropy implies $ S_{gen}(\infty) \geq S_{gen}(v*)$. Since the equation (\ref{C6}) holds for any $v* \geq 0$, we can write the same equation for some  wedge which is at $v{**}(y) \geq v*(y)$, i.e
\begin{equation}\label{C7}
S_{rel}(\Phi||\Omega_{HH})(v{**})= S_{gen}(\infty)-S_{gen}(v**)
\end{equation}
Since the wedge at $v{**}$ is contained in the wedge at $v*$, the monotonicity of Araki's Type III relative entropy for QFTs implies
\begin{equation}\label{C8}
S_{rel}(\Phi||\Omega_{HH})(v*)-S_{rel}(\Phi||\Omega_{HH})(v**)\geq 0.
\end{equation}
Using (\ref{C7}) and (\ref{C6}), we get
\begin{equation}\label{C9}
S_{gen}(v**) \geq S_{gen}(v*)
\end{equation}
for all $v{**}\geq v*\geq 0$. This is the local version of GSL that we wish to obtain. Since the relative entropy has been used, at every step, we are dealing with finite quantities. The computation will continue to hold in $AdS$ \footnote{A subtlety in the $AdS$ case is discussed in the next paragraph.}, with the only difference being that the partial Cauchy slice will be $\mathcal{H}^+(v*)$. The computation is also identical for the Rindler wedge; the only difference is that the cyclic separating state at late times will be the Minkowski vacuum, and the modular Hamiltonian in equation (\ref{C2}) is with respect to this vacuum. Furthermore, this technique can be simply extended to any spacetime having a Killing horizon like Kerr, modulo our assumption about the existence of a special Gaussian state on the horizon. Notice that this computation is fundamentally dependent on modular inclusion and the fact that the null translation is a symmetry on the horizon --- this mainly results in the local modular Hamiltonian for all wedges of type $B$ (wedges  whose future null boundary coincides with part of the future horizon). As we have seen, the JSS conjecture is true for any wedge with a boundary which overlaps with a part of the Killing horizon. The generalized entropy at each cut is equal to the entropy of the type II algebra of the wedge associated with that cut in the sense of JSS.

In an asymptotically $AdS$ spacetime, a question that can be asked is the boundary dual of this construction. Modular Hamiltonians on time bands in the boundary have been discussed in \cite{JLMS} (see also eq.(98) in \cite{TROH}). However, we have been informed by Prof. E. Witten that the modular Hamiltonian of a proper subregion in the boundary does not have a splitting into left and right parts. Thus, in the asymptotically $AdS$ case, we cannot merely use $H_{ADM}$ to implement a crossed product construction. It is clear that one has to add an observer in addition to $H_{ADM}$ and implement the crossed product with respect to the observer. If we add the observer, then the question is what is the boundary dual of the observer. This is a question which we hope to address in the future. The meaning of the observer has also been extensively discussed in \cite{KJA}.

Finally, we end this section with a note comparing this derivation with the proof of the GSL by Wall \cite{AW} who also used the monotonicity of the relative entropy. The type II crossed product construction provides a natural renormalization scheme which was an assumption in Wall's proof. Further, the relative entropy used in this section is Araki's type III relative entropy and all computations are done in modular theory.
\section{Discussion}
Crossed product constructions have proved to be very useful in renormalizing quantities such as one-sided modular Hamiltonians and associating an entropy with the algebra of field operators in subregions. However, so far, it has only been possible to obtain a weak form of a GSL in black hole spacetimes in crossed product constructions \cite{CPW}. This involves considering asymptotically $AdS$ black holes and proving that for a very large time gap between early and late times, the generalized entropy at late times is greater than at early times.
In this paper, we primarily show that a local version of the GSL, namely $\frac{dS_{gen}}{dv} \geq 0$, follows from crossed product constructions. The new ingredient is the application of recent results on entropy of the algebra of operators on subregions of general spacetimes by JSS \cite{KJA}. We discuss a slight generalization of the JSS construction in the case of asymptotic wedges where we explicitly introduce an observer and implement the crossed product with respect to the observer. This allows us to also obtain a GSL for asymptotically $AdS$ black holes, for which, as discussed in section V, we need to add an observer. Such extra degrees of freedom do not change our results, since these degrees of freedom cancel out in a single wedge when considering the relative entropy in the wedge.

We first use half-sided modular inclusions to obtain expressions for modular Hamiltonians for algebras of null-shifted wedges along the future horizons in maximally extended static black hole spacetimes. We also outline a similar computation for the horizon of the Kerr black hole. Then we apply the result of JSS to these wedges. The results of JSS rely on the conjecture that the Hamiltonian generating the flow of a specific vector field on the Cauchy slice is a modular Hamiltonian of some state. This conjecture is true in the setting to which we apply these results. \emph{It also allows us to interpret the generalized entropy at each cut on the horizon as the entropy of the algebra of the wedge associated with that cut} in the sense of JSS.  Further, we are able to compare relative entropy for two different subregions (specifically, wedges along the horizon) using the JSS results since the modular Hamiltonian used for the crossed product construction in both the wedges is \emph{for the same state}. How to compare, for example, algebra entropy of two different subregions of a spacetime in general in the JSS construction is an interesting open problem, with many potential applications.

\subsection{Modular Hamiltonians of deformed half-spaces:}
In Appendix B, we also compute modular Hamiltonians in a class of static spacetimes (including the Schwarzschild spacetime), which are modular Hamiltonians for the domains of dependence of deformed Cauchy slices of half-space using path integrals. The purpose is to check whether such a modular Hamiltonian, which is expected to be non-local, can be made local by adding two operators, one from the algebra, and one from the commutant (for the two-sided modular Hamiltonian) as surmised by JSS. We compute these modular Hamiltonians using the path integral method. The results produce for the two-sided Hamiltonian, apart from a local integral on a Cauchy surface, operators of the form $\int_{\mathcal{H}_{\pm}} T_{\pm \pm}\xi^{\pm}$ . Since perturbation theory requires $|\int_{\mathcal{H}_{\pm}} <T_{\pm \pm}>\xi^{\pm}|<1$, so it might be possible that there exists some operator $a$ in the algebra and an operator $b'$ in its commutant such that the difference of the deformed and undeformed modular Hamiltonians is $a+b'$. Along the way, we see that the averaged null energy condition (ANEC) also holds for null generators of the Cauchy horizon in the class of static spacetimes we have considered, which includes the Schwarzschild spacetime.
\section{Acknowledgements} MA acknowledges the Council of Scientific and Industrial Research (CSIR), Government of India for financial assistance. We thank Prof. E. Witten for a comment on the earlier version of our paper.
\appendix
\section*{Appendix}
\vspace{.3cm}
\section{Minkowski wedges}\label{app1}
\begin{figure}[h]
  \centering
  \includegraphics[width=0.80\textwidth]{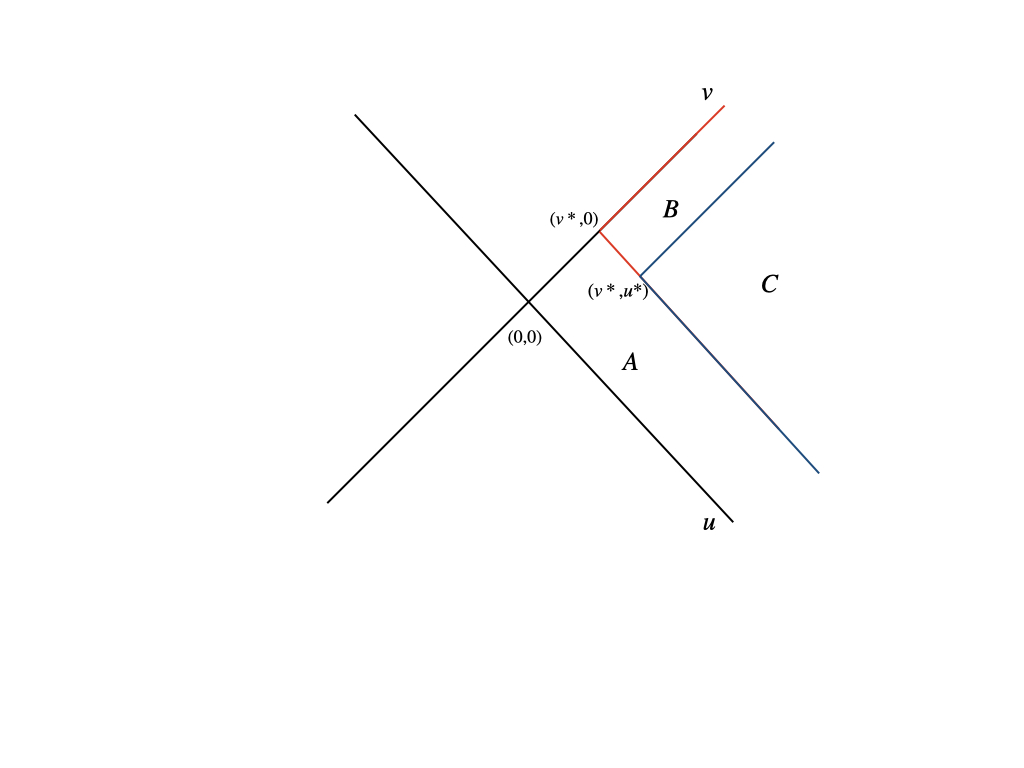}
  \caption{$A$,$B$ and $C$ are three Rindler wedges. $A$ is the Rindler wedge at the centre, $B$ is the wedge $A$ shifted along the null coordinate $v$ by $v*$ and $C$ is the wedge $B$ shifted along the null coordinate $u$ by $u*$. The coordinates in the diagram are the null coordinates and transverse coordinates are suppressed.}
  \label{fig:MW1}
\end{figure}

The objective of this section is to establish the relationship between the modular operator of the Rindler wedge \textbf{A}, whose null boundaries intersect at the origin of the Minkowski space, and another Rindler wedge \textbf{C}, which is contained within the wedge \textbf{A} and has no overlapping null boundaries with the wedge \textbf{A}, as shown in the Figure \ref{fig:MW1}. A second wedge, \textbf{B}, is introduced for computational purposes and for a subsequent section. Its null boundary overlaps with the part of the future null boundary of that of the wedge \textbf{A}.
Let $\mathcal{M_A}$, $\mathcal{M_B}$, and $\mathcal{M_C}$ be the von Neumann algebras associated with the wedges \textbf{A}, \textbf{B}, and \textbf{C} correspondingly, and let these algebras act on the Hilbert space $\mathcal{H}$. In the Minkowski spacetime, the Reeh-Schlieder theorem provides us a cyclic and separating state $\Omega$ (the Minkowski vacuum) for the von Neumann algebra of any proper subregion in the spacetime. In modular theory, we may define ($\Delta_A$, $J_A$), ($\Delta_B$, $J_B$), and ($\Delta_C$, $J_C$) as the modular operator and the modular conjugation associated with $(\mathcal{M_A},\Omega)$, $(\mathcal{M_B},\Omega)$, and $(\mathcal{M_C},\Omega)$, respectively.
\\
We will obtain the relationship between the modular operator of $\mathcal{M_A}$ and $\mathcal{M_C}$ in the following three steps:
\\
 i) We will prove that $\mathcal{M_B}$ is the positive modular inclusion of $\mathcal{M_A}$, and use its properties to derive the relationship between the modular Hamiltonians of $(\mathcal{M_A},\Omega)$ and $(\mathcal{M_B},\Omega)$. We will then demonstrate that $\Delta_B^{it}$ has a geometrical action on the wedge \textbf{B}.
\\
 ii) Following the same analysis as for the wedge $A$ and $B$, we will obtain the relation between the modular Hamiltonian of $(\mathcal{M_B},\Omega)$ and $(\mathcal{M_C},\Omega)$, showing that $\mathcal{M_C}$ is the negative half-sided modular inclusion of $\mathcal{M_B}$.
 \\
 iii) Using the previously obtained relation, we will obtain the relation between the modular Hamiltonian of $\mathcal{M_A}$ and $\mathcal{M_C}$.
 \\
 We have already defined modular inclusions in the main body of the paper, following \cite{HJB}.\vspace{5mm}
\\

\begin{figure}[h]
  \centering
  \includegraphics[width=0.80\textwidth]{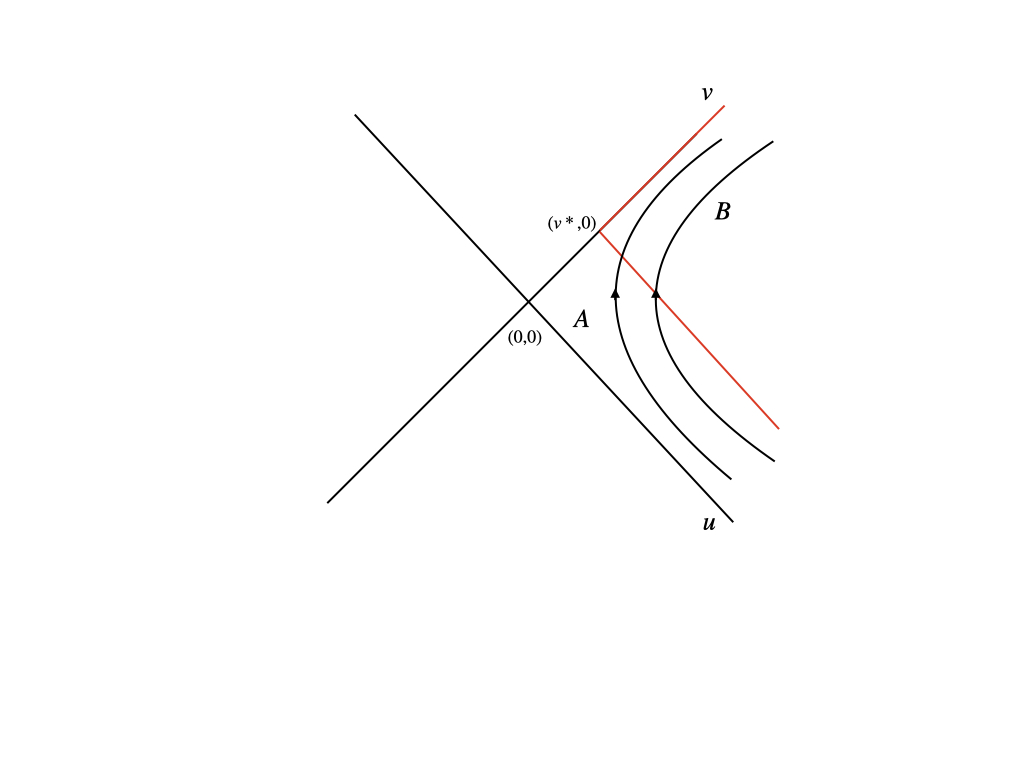}
  \caption{The figure represents boost integral curves which also represents the modular flow in the wedge $A$.}
  \label{fig:MW2}
\end{figure}
\textbf{Claim:} $\mathcal{M_B}$ is the positive modular inclusion of $(\mathcal{M_A}, \Omega)$.
\\Note that $\mathcal{M_B} \subset \mathcal{M_A}$. As previously stated, $\Omega$ is cyclic and separating for $\mathcal{M_B}$. According to the Bisognano Wichmann theorem \cite{BW}, $\Delta_A^{it}$ is the boost flow in the forward direction for the wedge $A$ when $t\leq0$. Thus, $\Delta_A^{it}$ has a geometrical action on the operators in $\mathcal{M_B}$, i.e. it moves the operators along integral curves of the boost Killing field as shown in Figure \ref{fig:MW2}. Because the boost is null on the Rindler horizon of wedge $A$ and timelike inside, the forward boost cannot take the local operator in $\mathcal{M_B}$ outside it. Therefore, $\Delta_A^{it} \mathcal{M_B} \Delta_A^{-it} \subset \mathcal{M_B}$ when $t\leq 0$. According to the definition of positive half-sided modular inclusion, $\mathcal{M_B}$ is a positive half-sided modular inclusion of $(\mathcal{M_A}, \Omega)$.
\\
For a more detailed study, let the vertices of the wedge $A$ and $B$  be separated by $v*$ along the null direction $v$, as shown in Figure \ref{fig:MW2}. Now, according to the results on modular inclusions, there exists a unitary $U(t)$ such that
\begin{equation}\label{A2}
\Delta_A^{-it}\Delta_B^{it}=U(e^{2\pi t}-1)
\end{equation}
where $U(t)= \exp[i \mathcal{E}_{v*}t]$ and $\mathcal{E}_{v*}$ is a positive operator. $U(t)$ can be thought of as an operator that translates the wedge $B$ in null direction $v$. Further, $\mathcal{E}_{v*}$ can be written as $v*$ times generator of null translation along the $v$. We can write the above equation as
\begin{equation}{\label{A3}}
\exp[-it\log[\Delta_A]]\exp[it \log[\Delta_B]]=\exp[i(e^{2\pi t}-1)\mathcal{E}_{v*}]
\end{equation}
Now, differentiating the above equation with respect to $t$ and evaluating it at $t=0$ gives
\begin{equation}{\label{A4}}
\log[\Delta_A]-\log[\Delta_B]= -2\pi \mathcal{E}_{v*}
\end{equation}
Now, if we can define the modular Hamiltonian as $K=-\log[\Delta]$, then
\begin{equation}{\label{A5}}
K_B= K_A-2\pi \mathcal{E}_{v*}.
\end{equation}
We want to emphasize that the above result is true even if $v*$ depends on the transverse coordinate. However, if $v*$ depends on the transverse coordinate, the modular flow generated by it will not have a local action on the wedge $B$ but it will have local action along the null boundary (horizon) associated with the wedge. Nevertheless, for $v*$ independent of the transverse coordinate, the modular flow is local and that is what we will assume for rest of this section.
\\
\textbf{Claim:} $K_B$ is a boost generator associated with wedge $B$.
\\
 To show that $K_B$ is a boost generator associated with wedge $B$, we use the fact that we can use Theorem $1$. Differentiating the  condition ($h$) in the theorem $1$ first with respect to $t$  and evaluating at $t=0$ and then with respect to $s$ and evaluating it at $s=0$, we will get
 \begin{equation}{\label{A6}}
 [iK_A,i\mathcal{E}_{v*}]=2\pi i\mathcal{E}_{v*}.
 \end{equation}
 From (\ref{A3}), we can write
 \begin{equation}{\label{A7}}
 \exp[-it K_B]=\exp[-itK_A]\exp[i(e^{2\pi t}-1)\mathcal{E}_{v*}]
 \end{equation}
 Now there is a well-known theorem which we will just use here.
 \\
 \textbf{Theorem 2:} If $[X,Y]=sY$, where $s \in \mathbb{C}$  and $s\neq 2\pi i n$  then\\
 \begin{equation}\label{A8}
 \exp[Y]\exp[-X]\exp[-Y]=\exp[-X]\exp[(\exp[s]-1)Y]
 \end{equation}
 Now choose $X=iK_A t$ and $Y=i\mathcal{E}_{v*}$. Then using (\ref{A6}), one can identify $s=2\pi t$. Since $t \in \mathbb{R}$, we can apply the theorem. This gives
 \begin{equation}\label{A9}
 \exp[-itK_A]\exp[i(e^{2\pi t}-1)\mathcal{E}_{v*}]= \exp[i\mathcal{E}_{v*}]\exp[-iK_A t]\exp[-i\mathcal{E}_{v*}]
 \end{equation}
 Now putting (\ref{A9}) back in (\ref{A7}), we get
  \begin{equation}\label{A10}
  \Delta_B^{it}=U(1)\Delta_A^{it}U(-1)
  \end{equation}
  So this is a null translated boost, which can still be thought of as a boost but this time associated with the wedge $B$. Furthermore, (\ref{A5}) and (\ref{A10}) establish its local and geometrical nature. There is another way to get (\ref{A10}), because $\mathcal{M_B}=U(1) \mathcal{M_A} U(-1)$, and $U$ is an $\Omega$ preserving unitary. It is straightforward to verify that the Tomita operator for $B$ is $S_B= U(1) S_A U(-1)$.
  \\
\begin{equation}\label{A11}
  U(1) S_A U(-1) \Big(U(1) a U(-1)\Big) \Omega=U(1) a^{\dag} U(-1)\Omega
\end{equation}
For each $a \in \mathcal{M_A}$, $U(1)aU(-1) \in \mathcal{M_B}$. Now, using the definition of the modular operator $\Delta_B=S_B^{\dag}S_B= U(1)\Delta_AU(-1)$. Further, using the spectral theorem for operators and the fact that $U(1)$ is unitary, we will get the equation in (\ref{A10}).
\vspace{3mm}
 \\
 \begin{figure}[h]
  \centering
  \includegraphics[width=0.80\textwidth]{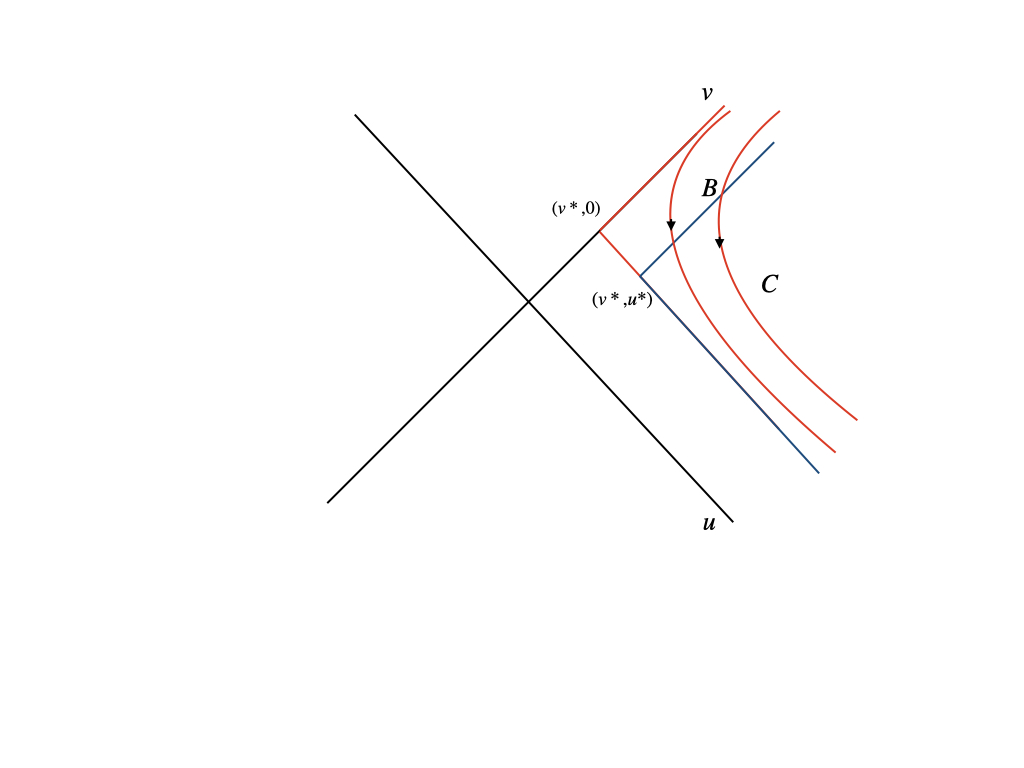}
  \caption{The figure represents boost integral curves associated with the wedge $B$ which also represents the modular flow in the wedge $B$.}
  \label{fig:MW3}
\end{figure}
 We can now do the same with wedges $B$ and $C$. As we already know, $\Omega$ is cyclic and separating for $\mathcal{M_C}$.  $\Delta_B^{it}$ is the boost associated with the wedge $B$ and the past Rindler horizon of the wedge $C$ overlaps with the portion of the past horizon of the wedge $B$, as shown in Figure \ref{fig:MW3}. For $t \geq 0$, the boost $\Delta_B^{it}$ maps the wedge $C$ into itself. Thus, $\mathcal{M_C}$ is a negative inclusion of $\mathcal{M_B}$. Following the steps of previous analysis, using Theorem $1$ we can write
 \begin{equation}\label{A12}
\Delta_B^{-it}\Delta_C^{it}=V(1-e^{-2\pi t})
\end{equation}
where $V(t)= \exp[i \mathcal{E}_{u*}t]$  and $\mathcal{E}_{u*}$ is a positive operator. $V(t)$ can be thought of as the operator that translates wedge $B$ in the null direction $u$. Further $\mathcal{E}_{u*}$ can be written as $u*$ times generator of null translation along $u$. We obtain
\begin{equation}\label{A13}
K_C= K_B-2\pi \mathcal{E}_{u*}
\end{equation}
and
\begin{equation}\label{A14}
 \Delta_C^{it}=V(-1)\Delta_B^{it}V(1).
\end{equation}
The modular Hamiltonians of the algebra and the algebra related by the modular inclusion will differ by the generator of a one-parameter unitary group. If the algebras are wedge algebras with inclusion as a null translated wedge, then the generator that connects the two modular Hamiltonians is a null translation generator. Now we may express the modular Hamiltonian of the wedge $C$ in terms of the modular Hamiltonian of $A$. Using (\ref{A13} and (\ref{A5}), we obtain
\begin{equation}\label{A15}
K_C= K_A-2\pi \mathcal{E}_{v*}-2\pi \mathcal{E}_{u*}.
\end{equation}
Further using (\ref{A14}) and (\ref{A10}), we can write the modular flow
\begin{equation}\label{A16}
\Delta_C^{it}=V(-1)U(1)\Delta_A^{it}U(-1)V(1).
\end{equation}
Since translations in Minkowski spacetime commute, we can define the unitary $W(s;a,b)=exp{[2\pi i(a \mathcal{E}_{v*}+b \mathcal{E}_{u*})s]}$ and the equation can be written as
\begin{equation}\label{A17}
\Delta_C^{it}=W(1;1,-1)\Delta_A^{it}W(-1;1,-1)
\end{equation}
This is true for any $v*$ and $u*$, therefore we have obtained a relation between the modular Hamiltonian of any wedge that can be reached via a series of null translations and the wedge at the origin. It is crucial to note that because null translation is a global isometry of spacetime, the resulting modular Hamiltonians for $B$ and $C$ are conserved and may be represented as a local integral on the Cauchy surface.  This is the simplest example of a local modular Hamiltonian. The result (\ref{A5}) is valid even if $v*$ depends on the transverse coordinate. The only difference us that the null translation that maps two wedges would not be a symmetry and therefore the resultant modular Hamiltonian may not have local action everywhere inside the wedge $B$. But it will be local on the horizon, since the null translation depending on the transverse coordinate is still a symmetry on the horizon.
\section{Modular Hamiltonian of deformed half-spaces in general spacetimes}

In this section, we depart somewhat from the techniques of the previous sections and consider (one-sided) modular Hamiltonians (for the vacuum state) computed using path integrals rather than Tomita-Takesaki theory. The issue then is that the one-sided modular Hamiltonian may be formally infinite, however we will assume that this can be renormalized, since we will finally be interested in two-sided modular Hamiltonians \footnote{The one-sided modular Hamiltonian is well defined as a Hermitian form on a dense set of states.}. The purpose is to compute some examples of general modular Hamiltonians to confirm/check the expectation of JSS \cite{KJA} that a non-local modular Hamiltonian $H$ of some state $| \Psi >$ may be made local by subtracting off an element of the algebra $a$ and an element of its commutant, $b'$ such that $H' = H - a - b'$ is local. By local, we mean a local integral over some Cauchy slice. Then the converse of the cocycle derivative theorem implies that $H'$ is the modular Hamiltonian of some other state $| \Psi_{ab'} >$.

The expectation that the modular Hamiltonian can be made local is crucial to the conjecture of JSS that $H^g_\xi$ is the modular Hamiltonian for some state on the algebra $\mathcal{A}_{QFT}$. In order to check this, we need modular Hamiltonians at our disposal, which are hard to calculate in situations without a lot of symmetry. In cases where, for example, there is Killing symmetry, the modular Hamiltonian for the vacuum can be computed in most situations and is a local expression on a Cauchy slice. However, other than these examples, the modular Hamiltonian will in general, be non-local and will not generate a geometric flow on the spacetime.

\begin{figure}[h]
  \centering
  \includegraphics[width=1.0\textwidth]{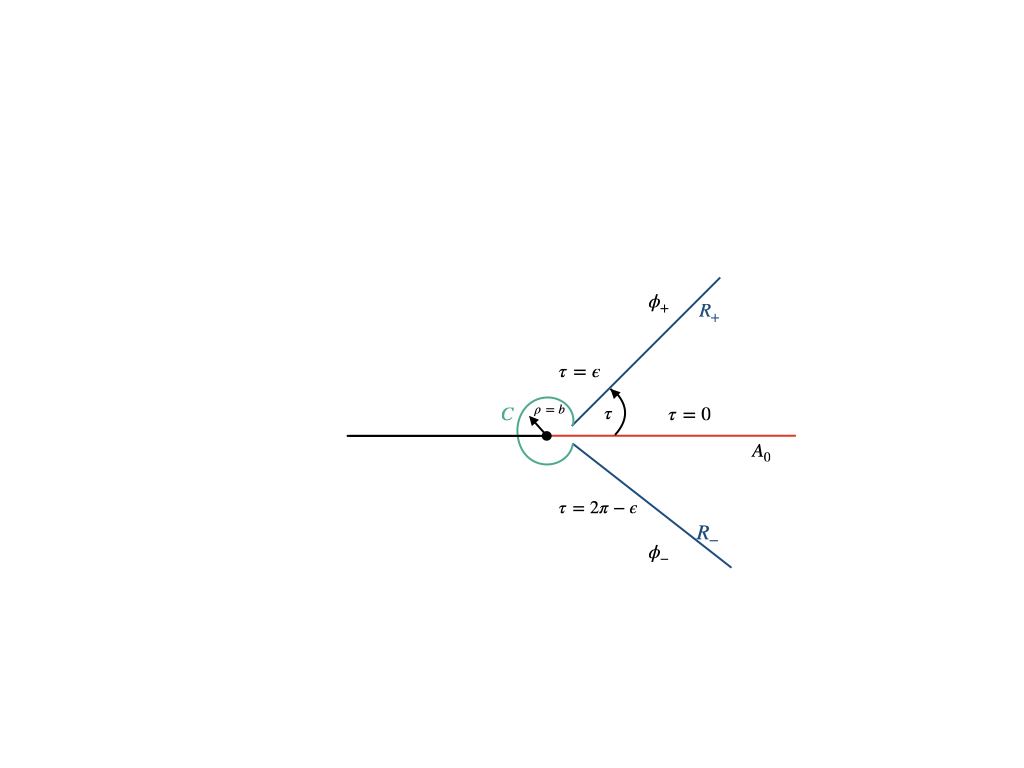}
  \caption{The red line represents branch cut in the $(\rho,\tau)$ plane, the blue line $R_\pm$ will give the contribution from the branch cut and C is the $\rho=b$ surface which will give the contribution from the entangling surface as $b\rightarrow 0$.}
  \label{fig:DMH2}
  \end{figure}

The simplest example is to consider the half-space in Minkowski spacetime, whose domain of dependence is the Rindler wedge. We know the modular Hamiltonian --- it is the generator of a boost and therefore can be written as a local integral on the $t=0$ surface. Consider an arbitrary deformation of the $t=0$ surface by a small amount which, in particular, perturbs the entangling surface itself.  As is well-known, the modular Hamiltonian on this surface will be non-local, but we can do a perturbative expansion about the modular Hamiltonian associated with the half-space at $t=0$ and obtain a relation between them. This was obtained by Faulkner, Leigh, Parrikar and Wang (FLPW) in \cite{TROH}  and later by Balakrishnan and Parrikar (BP)\cite{SO} for Minkowski spacetime and in the paper \cite{FR} by Rosso for $AdS_2 \times S^{d-2}$. This technique gives, at least perturbatively, the modular Hamiltonian for more general wedges in the spacetime.
\\
We would like to see if the perturbative correction to the half-space modular Hamiltonian can be made local as surmised by JSS by subtracting off a piece $a$ from the modular Hamiltonian (and a piece $b'$ from the commutant when one considers the two-sided modular Hamiltonian).
But first, we want to apply the technique of FLPW to a more general class of spacetimes and find the relation between the half-space modular operator and the deformed one. We will work with the Wick-rotated Euclidean metric. The class of metrics we are interested in is the class of Wick-rotated metrics with the form
\begin{equation}\label{D1}
ds^2=g_{\mu \nu}dX^\mu dX^\nu=\exp[\Omega(\rho)]\Big(d\rho^2 + f(\rho) d\tau^2\Big)+ h_{ab}(\rho^2,\vec{x}) dx^a dx^b
\end{equation}
where $\Omega(\rho)$ is a smooth function for all $\rho$ and $f(\rho)$ is a positive function, which goes like $\kappa^2\rho^2$ for small $\rho$ where $\kappa$ is some constant. We will also assume $h_{ab}$ is a Riemannian metric which is smooth at $\rho=0$.  $\tau$  is Euclidean time with periodicity $\beta=\frac{2\pi}{\kappa}$. Note, the metric components are independent of $\tau$. The periodicity ensures that there is no conical singularity. Since the metric is independent of $\tau$, the metric after the Wick rotation corresponds to a static Lorentzian metric. Furthermore, the metric is degenerate at $\rho=0$ indicating the existence of a horizon associated with these coordinates. After the Wick rotation, one can analytically continue the coordinates to obtain its maximal extension.
The metric ansatz accommodates many interesting cases. The metric is conformally Rindler if $f(\rho)=\rho^2$ and transverse metric is flat, Rindler if we also have $\Omega(\rho)=0$.  It is $AdS_2 \times M_{transverse}$  when $f(\rho)=\sinh^2\rho$ and $\Omega(\rho)=0$. The Schwarzschild metric can also be put in this form.
\begin{figure}[h]
  \centering
  \includegraphics[width=1.0\textwidth]{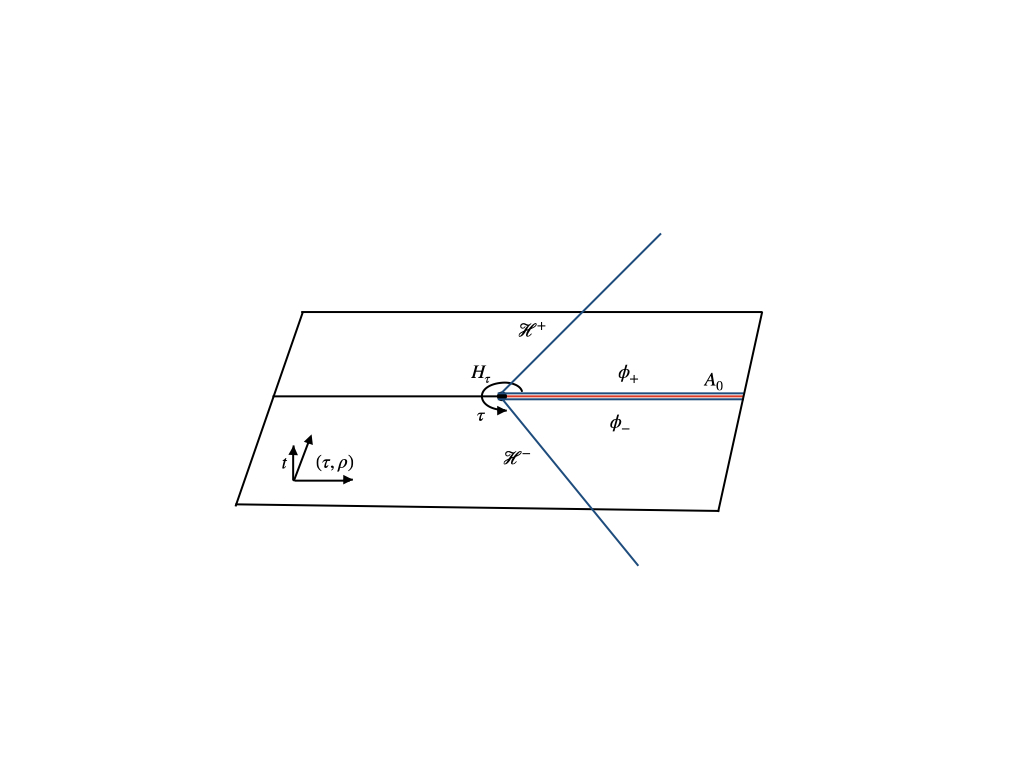}
  \caption{The red line represents the surface $A_0$ on which density matrix is computed in $(\tau,\rho)$ plane. $H_{\tau}$ is the generator that maps $\phi_{+}$ to $\phi_{-}$ on $(\tau,\rho)$. $t$ in the figure is Lorentzian time and $\mathcal{H}^{\pm}$ are the Cauchy (and Killing) horizons of the Cauchy slice.}
  \label{fig:DMH3}
  \end{figure}
\\
 We are interested in computing the modular Hamiltonian for some Cauchy slice which is not a half space in these class of spacetimes. Consider a QFT in the above spacetime. It is well-known how to compute the density matrix of the vacuum state using the Euclidean path integral on any generic spacetime with the time translation symmetry. The spacetime should have a well-defined Wick rotation, and the metric should be smooth everywhere after the rotation. The density matrix for the vacuum state on an arbitrary surface is obtained in  \cite{GILD} and is non-local, as expected.  The nonlocality arises from the fact that the generator that maps the configuration of fields above and below the surface of interest on which the density matrix has to be computed is not a symmetry. In our case, we want to compute the density matrix for a surface which is a small deformation of the $\tau=0$ surface. But, first let us compute the density matrix for the $\tau=0$ surface as shown in Figure \ref{fig:DMH3}. Since $\tau$ translation is an isometry of the spacetime, the result will be local. In general, we will get
\begin{equation}\label{D2}
\rho_{A_0,g}= \exp[-\beta H_{\tau}]
\end{equation}
$\rho_{A_0,g}$ represents the density matrix \footnote{$g$ in $\rho_{A_0,g}$ is just to emphasize that it is defined in the metric g.}. $H_{\tau}$ is a generator of $\tau$ translation. $H_{\tau}$ can be written in terms of the QFT stress tensor and vector field $\partial_{\tau}$ \cite{GILD}. Now the one-sided modular Hamiltonian of the vacuum can be obtained using  $K_{A_0,g}=-\log \rho_{A_0,g}=\beta H_{\tau}$.  This is the modular Hamiltonian associated with the domain of dependence $D(A_0)$. The modular Hamiltonian derived by Sewell for the right exterior of the Schwarzschild black hole in \cite{GLS} is of this form. In fact, the class of (Lorentzian) metrics we consider in this section are exactly of the form assumed by Sewell. So we can simply use Sewell's computation for getting the modular Hamiltonian. Now, we are interested in the modular Hamiltonian of the deformed region $D(A)$ associated with the Cauchy surface $A$, which is obtained via a small diffeomorphism of $A_0$. Let $X^{\mu}=X'^{\mu}-\zeta(\rho,\vec{x})$ be the diffeomorphism which maps $A_0$ to $A$, where $\zeta$ is generator of this infinitesimal diffeomorphism. Note, we also assume it to be independent of $\tau$. We will assume that $\zeta$ is non-vanishing and smooth at $\rho=0$. The unitary operator that implements this diffeomorphism on the Hilbert space  is \cite{TROH,FR},
\begin{equation}\label{D3}
U=\exp[\int_{\tau=0}d\Sigma^{\mu}T_{\mu \nu}\zeta^{\nu}]
\end{equation}
where $T_{\mu \nu}$ is the stress tensor. Note that there is no $i$ in the above equation, since we are working with the Euclidean theory. By applying a general identity for computing the derivative of the $\log$ of an operator, FLPW \cite{TROH} have shown that
\begin{equation}\label{D4}
K_{A,g}=K_{A_0,g}+[K_{A_0,g},\delta U]+\delta K_{A_0,g}+O(\zeta^2)
\end{equation}
where $K_{A,g}$ is the modular Hamiltonian of the deformed surface, $\delta U$ is a linear order term in $\zeta$ when $U$ is expanded in $\zeta$ and $\delta K_{A_0,g}$ is,
\begin{equation}\label{D5}
\delta K_{A_0,g}= \int_{-\infty+i\alpha}^{\infty+i\alpha} \frac{dz}{4\sinh^2{z/2}}\int_{\partial M_E } dS^{\mu}\rho_{A_0,g}^{-\frac{iz}{2\pi}}T_{\mu \nu} \zeta^{\nu}\rho_{A_0,g}^{\frac{iz}{2\pi}}
\end{equation}
where $\alpha \in (0,2\pi)$ is a free parameter, $z=s+i\alpha$ where $s\in \mathbb{R}$ and $\partial M_E$ is the boundary of the Euclidean manifold $M_E$ we are working on. Here we have put $\beta=2\pi$, so that we do not have to track it at each step, but one can introduce it and it will just change the answer by scaling of $\beta$\footnote{To do the computation with $\beta$, let $\rho_{A_0,g}^{\frac{iz}{2\pi}} \rightarrow \rho_{A_0,g}^{\frac{iz}{\beta}}$ in the equation (\ref{D5}). }. The above integral does not get a contribution from the conformal boundary\footnote{We can take $\zeta$ be non-vanishing only for very small $\rho$ and at $\tau=0$, the contribution coming from the spatial boundary dies off due to appropriate fall-off of the stress tensor.}. The only contributions we get are from the branch cut and $C$ , i.e $\partial M_E= C \cup R_+ \cup R_-$ as shown in the Figure \ref{fig:DMH2}, for more details, see \cite{TROH,FR}.
 Further, we can split the contribution as coming from $C$ and $R_+\cup R_-$,
 \begin{equation}\label{D6}
 \delta K_{A_0,g}= \delta K_{A_0,g,C}+\delta K_{A_0,g,R_+\cup R_-}
 \end{equation}
 \textbf{Contribution from C:}
 \\
 We are interested in computing the contribution of $C$ as $b\rightarrow 0$. Since $\rho$ is very small on the contour $C$, we can work we the metric
 \begin{equation}\label{D7}
 ds^2=\exp[\Omega(\rho)]\Big(d\rho^2 + \rho^2 d\tau^2\Big)+ h_{ab}(\rho^2,\vec{x}) dx^a dx^b
 \end{equation}
 It will be more convenient to work in the following variable,
 \begin{equation}\label{D8}
 x_{\pm}=\rho\exp[\mp i\tau]
 \end{equation}
 It is straightforward to show
 \begin{equation}
 ds^2=\exp[\Omega(\rho)]\Big(dx_+ dx_-\Big)+ h_{ab}(\rho^2,\vec{x}) dx^a dx^b ,
 \end{equation}
 the translation in $\tau$ is scaling in $x_{\pm}$, i.e $\tau \rightarrow \tau +iz $ becomes $ x_{\pm} \rightarrow e^{\pm z} x_{\pm}$. For computing $\delta K_{A_0,g,C}$ , we need the inward unit normal $n^\mu$ to $C$, which can easily be obtained,
  \begin{equation}\label{D9}
  n^\mu =-\exp[\Omega(b)/2](e^{-i\tau}\delta^{\mu}_+ + e^{i\tau}\delta^{\mu}_-)
  \end{equation}
  We know that $\rho_{A_0}^{-\frac{iz}{2\pi}}$ generates the diffeomorphism $x_{\pm} \rightarrow  \bar{x}_\pm=e^{\pm z} x_{\pm}$. We can write
  \begin{equation}\label{D10}
\rho_{A_0,g}^{-\frac{iz}{2\pi}}T_{\mu \nu}(x_{\pm}, \vec{x}) \rho_{A_0,g}^{\frac{iz}{2\pi}}= \frac{\partial\bar{x}^{\gamma}}{\partial x^\mu}\frac{\partial \bar{x}^{\beta}}{\partial x^\nu} T_{\gamma \beta}(\bar{x}_{\pm}, \vec{x})
  \end{equation}
Since $\alpha$ in the limits of integration in (\ref{D5}) is a free parameter, we will work with the choice $\alpha= \tau$. Using (\ref{D9}) and (\ref{D10}), we can show $B=n^\mu \zeta^{\nu}\rho_{A_0,g}^{-\frac{iz}{2\pi}}T_{\mu \nu} \rho_{A_0,g}^{\frac{iz}{2\pi}}\Big|_C$ is
 \begin{equation}\label{D11}
B=-e^{-\Omega(b)/2}\Big(\zeta^+ e^{s+i\tau}(T_{++}e^s + T_{+-}e^{-s})+\zeta^- e^{-(s+i\tau)}(T_{--}e^{-s} + T_{-+}e^{s})+\zeta^a(T_{+a}e^s + T_{-a}e^{-s})\Big).
\end{equation}
Since
\begin{equation}\label{D12}
\delta K_{A_0,g,C}= \lim_{b\rightarrow 0}\int \sqrt{h}dx^{d-2}\int_{-\infty}^{\infty}\frac{ds}{4\sinh^2(\frac{s+i\tau}{2})}\int_0^{2\pi} b e^{\Omega(b)/2} d\tau B ,
\end{equation}
note that $e^{\Omega(b)/2}$ in the equation cancels the $e^{-\Omega(b)/2}$ in $B$. Therefore the contribution from $C$ is independent of the conformal factor in the metric (\ref{D1}). We can split the above equation in the components of $\zeta$, as
\begin{equation}\label{D13}
\delta K_{A_0,g,C}=\delta K_{A_0,g,+}+\delta K_{A_0,g,-}+\delta K_{A_0,g,a}
\end{equation}
where
\begin{equation}\label{D13A}
\delta K_{A_0,g,\pm}=-\lim_{b\rightarrow 0} \int \sqrt{h}dx^{d-2} \zeta^{\pm}(b,\vec{x})\int_{-\infty}^{\infty}dsI_{\pm}(s)b\Big(T_{\pm \pm}e^{\pm s} + T_{+-}e^{\mp s}\Big)
\end{equation}
where
\begin{equation}\label{D14}
I_{\pm}(s)=\int_0^{2\pi}d\tau \frac{e^{\pm(s+i\tau)}}{4\sinh^2(\frac{s+i\tau}{2})}
\end{equation}
and
\begin{equation}\label{D15}
\delta K_{A_0,g,a}=-\lim_{b\rightarrow 0} \int \sqrt{h}dx^{d-2} \zeta^{a}(b,\vec{x})\int_{-\infty}^{\infty}dsI_{a}(s)b\Big(T_{+a}e^{ s} + T_{-a}e^{- s}\Big)
\end{equation}
where
\begin{equation}\label{D16}
I_{a}(s)=\int_0^{2\pi}d\tau \frac{1}{4\sinh^2(\frac{s+i\tau}{2})}
\end{equation}
Now to compute $I$, we will use the well-known identity
\begin{equation}\label{D15A}
\bar{I}_{J}=\oint d\omega \frac{\omega^J }{\omega-e^{-s}}= 2\pi i \Theta_J e^{-Js}
\end{equation}
where $J$ is integer and
\[
\Theta_J =\begin{cases}
  \Theta(s) &   J \geq0  \\
 -\Theta(-s) & J<0
\end{cases}
\]
where $\Theta$ is the step function. It is easy to show that $I_a=i \frac{\partial\bar{I}_{0}}{\partial s}= 2\pi \delta(s)$ and $I_{\pm}=i\frac{\partial\bar{I}_{\pm 1}}{\partial s}e^{\pm s}$, where
\begin{equation}\label{D16A}
\frac{\partial\bar{I}_{\pm 1}}{\partial s}= 2\pi i (\delta(s)e^{\mp s}-\pm e^{\mp s}\Theta_{\pm 1})
\end{equation}
We eventually want to compute the integrals (\ref{D13}) and (\ref{D15}) as $b\rightarrow 0$, but they are non-vanishing only if $b\rightarrow 0$ is compensated by $s\rightarrow \pm \infty$. Since $I_a=2\pi \delta(s)$,  $\delta K_{A_0,g,a}$ vanishes as $b\rightarrow 0$. Similarly, $\delta(s)$ terms in $I_{\pm}$ do not contribute, leaving only $\Theta$ terms to contribute. Notice that the term with $T_{+-}$ does not contribute. Therefore, we are left with
\begin{equation}\label{D17}
\delta K_{A_0,g,\pm}=- 2\pi \int \sqrt{h}dx^{d-2} \zeta^{\pm}(\vec{x})\int_{0}^{\pm\infty}dx_{L\pm}T_{\pm \pm}(x_{L\pm},\vec{x})
\end{equation}
where $x_{L \pm}=be^{\pm s}$ are Lorentzian null coordinates. Therefore
\begin{equation}\label{D18}
\delta K_{A_0,g,C}= -2\pi \int \sqrt{h}dx^{d-2} \Big(\zeta^{+}(\vec{x})\int_{0}^{\infty}dx_{L +}T_{+ +}(x_{L\pm},\vec{x})+ \zeta^{-}(\vec{x})\int_{0}^{-\infty}dx_{L -}T_{- -}(x_{L \pm},\vec{x})\Big)
\end{equation}
\textbf{Contribution from $R_+ \cup R_-$:}
Let the unit normal to the $R_\pm$ be denoted by $n_{\pm}$. This is given by
\begin{equation}\label{D19}
n_\pm^{\mu}= \mp \frac{\delta^\mu_\tau}{\sqrt{e^{\Omega(\rho)}f(\rho)}}
\end{equation}
and the metric induced on this surface is
\begin{equation}\label{D20}
ds^2|_{R_{\pm}}=\exp[\Omega(\rho)]d\rho^2 + h_{ab}(\rho^2,\vec{x}) dx^a dx^b
\end{equation}
In computing the contribution from $R_{\pm}$, we will choose the free parameter $\alpha=\epsilon$ for $R_+$ and $\alpha=2\pi -\epsilon$  for $R_-$. $\rho_{A_0,g}^{-iz}$ generates $\tau$ translation. Using (\ref{D10}), we can write
\begin{equation}\label{D21}
\delta K_{A_0,g,R_\pm}=\int \sqrt{h}dx^{d-2} \int_b^{\infty} e^{\Omega(\rho)/2}d\rho \zeta^\nu n_{\pm}^\mu\int_{-\infty}^{\infty} \frac{ds}{4\sinh^2\frac{(s \pm i\epsilon)}{2}}\rho_{A_0,g}^{-is}T_{\mu\nu}(0,\rho,\vec{x})\rho_{A_0,g}^{is}
\end{equation}
since the contours for $R_\pm$ are oriented oppositely and $n^\mu_-=-n^\mu_+$. One can close the contour and from residue theorem the contribution comes only from the double pole at $s=0$. Further, one can show\footnote{For the residue at double pole, we have to compute the first derivative of the non-singular part at $s=0$ ,and will result in the commutator. }
\begin{equation}\label{D22}
\delta K_{A_0,g,R_-\cup R_+}= [\delta U,K_{A_0,g}] .
\end{equation}
Using (\ref{D22}), (\ref{D18})  and (\ref{D6}), we can obtain $\delta K_{A_0,g}$. Further, putting in (\ref{D4}), we get
\begin{equation}\label{D23}
K_{A,g}=K_{A_0,g}-2\pi \int \sqrt{h}dx^{d-2} \Big(\zeta^{+}(\vec{x})\int_{0}^{\infty}dx_{L+}T_{+ +}+ \zeta^{-}(\vec{x})\int_{0}^{-\infty}dx_{L-}T_{- -}\Big)+O(\zeta^2)
\end{equation}
 We have obtained the modular Hamiltonian of the deformed surface at leading order in the deformation field using the FLPW technique. $K_{A_0,g}$ is a local expression $$\int_{A_0} d\Sigma^{\mu} V^{\nu} T_{\mu \nu}$$ where $V$ is the Killing vector field. It is conserved (from the fact that the energy momentum tensor is conserved and that $V$ is Killing). Therefore, we can also evaluate it on a Cauchy slice that consists of the Killing horizon $\mathcal{H}^{+}$ and $\mathcal{I}^{+}$ (for asymptotically flat spacetimes). Then we can attempt to combine the second integral on the right in (\ref{D23}) which is also an integral on $\mathcal{H}^{+}$ with the contribution to $K_{A_0,g}$ from $\mathcal{H}^{+}$. Indeed, in a situation where $\zeta^{-}(\vec{x}) = 0$, we can do the computation to higher orders, and all of these will be integrals on $\mathcal{H}^{+}$. Presumably, they can be resummed \cite{SO} to get the one-sided modular Hamiltonian for a null translated wedge along $\mathcal{H}^{+}$ exactly as in the previous sections. But when $\zeta^{-}(\vec{x}) \neq  0$, we will not be able to combine the second integral in this manner. This is the signature, at leading order in the deformation field, that the modular Hamiltonian is really non-local. It can be made more explicit at higher orders in the deformation field, where the non-local products of the stress tensor will appear. This can be seen even in the  Minkowski spacetime, for example, in the equation (3.21) in \cite{SO}. We note also that the integrals are over the horizons of the domain of dependence of the slice $A_0$ and differ from similar integrals on the horizon of the domain of dependence of $A$ only at quadratic order in the deformation field.
 \\
 \begin{figure}[h]
  \centering
  \includegraphics[width=0.90\textwidth]{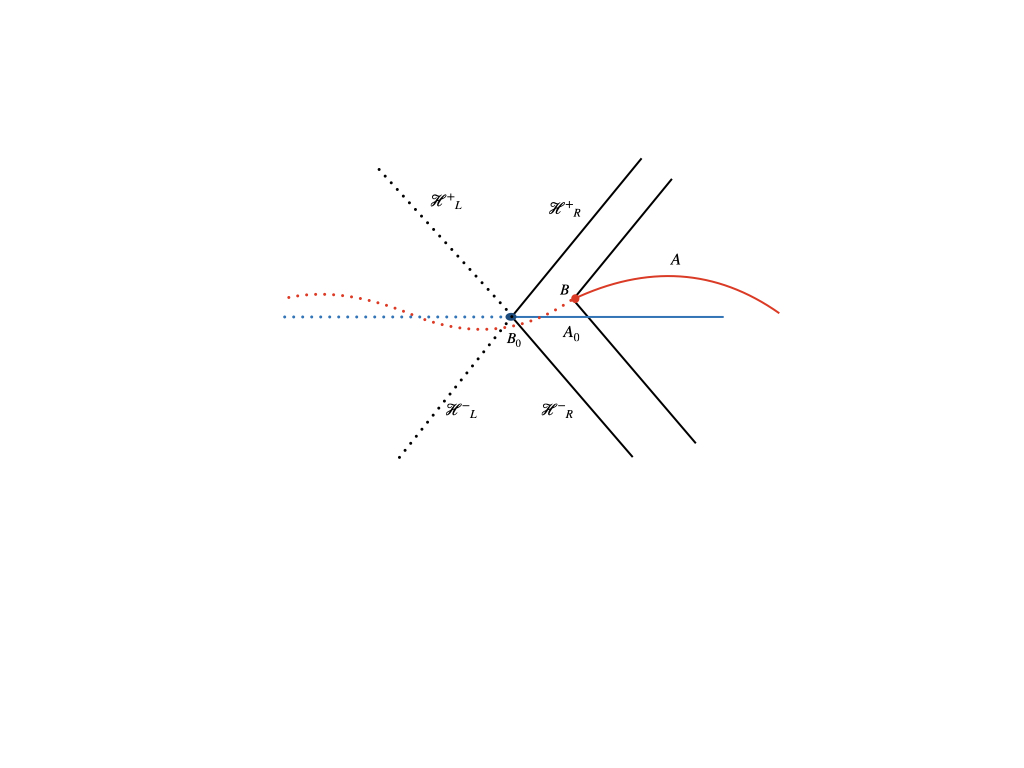}
  \caption{In this diagram $A_0$ is the undeformed partial Cauchy slice of the half space and $A$ is the deformed partial Cauchy slice. $B_0$  represents bifurcation surface associated with wedge with $A_0$  being Cauchy slice where $B$ represents bifurcation surface associated with the wedge with $A$ being Cauchy slice. }
  \label{fig:Cauchy}
\end{figure}
 Since the integrands in the terms in $K_{A,g}- K_{A_0,g}$ depend on the quantum fields due to the energy-momentum tensor, we can think of these terms as affiliated to the right exterior algebra (at linear order in the deformation field) and for the two-sided modular Hamiltonian, we have an identical term in the left half space affiliated to the commutant. Here by affiliated, we mean affiliated to the algebra or its commutant as a Hermitian form on some Hilbert space $\mathcal{H}$ \cite{speranza} \footnote{ See also Corollary 2.12 in \cite{sorce1} for operators affiliated to a von Neumann algebra.}. The above computation is a perturbative computation in the deformation field, and it only makes sense in the regime of perturbation theory. It restricts the state in which one can compute the expectation value of the deformed modular Hamiltonian. For example, if we work with states such that $|\int_{\mathcal{H}_{\pm}} <T_{\pm \pm}>\xi^{\pm}|<1$, this presumably can always be met by appropriately choosing the magnitude of the deformation field. It tells us that maybe we can think of the deformation field as a smoothing function. Notice that the difference of the two-sided modular Hamiltonians under the deformation depends on precisely $\int_{\mathcal{H}_{\pm}} T_{\pm \pm}\xi^{\pm}$ types of quantities coming from the wedge and its commutant. So, it might be possible that there exists some $a$ in the algebra and $b'$ in its commutant such that the difference of the modular Hamiltonians is $a+b'$ in the states where $|\int_{\mathcal{H}_{\pm}} <T_{\pm \pm}>\xi^{\pm}|<1$.
 We also see that these terms have support near the original entangling surface since the deformation actually changes the entangling surface --- see Figure (\ref{fig:Cauchy}). The distinction between the horizon of the original wedge and the deformed wedge only comes at quadratic order in the deformation.
This result is suggestive of the statement of JSS in \cite{KJA}, that a non-local two-sided modular Hamiltonian can be written as the sum of a local modular Hamiltonian of some other state and some operator from the algebra plus one from the commutant which makes the full operator non-local.

The construction of FLPW \cite{TROH} can be done for higher order terms in the deformation. It is expected that this can be resummed for null deformations given only by $\zeta^+(\vec{x})$ and one will get  \footnote{It seems that if one works with Gaussian null coordinates, a computation similar to \cite{SO} will go through.},
\begin{equation}\label{D24}
K_{A,g}= 2\pi \int \sqrt{h}dx^{d-2}\int_{\zeta{(\vec{x})}}^{\infty} (x_{L+} - \zeta^{+}(\vec{x}))T_{+ +} dx_{L+}.
\end{equation}
We have already seen this in previous sections for the Schwarzschild black hole, using HSMI. Translation symmetry in $\tau$ corresponds to a boost in null coordinates. Null translation at the horizon is a symmetry for the Schwarzschild metric. Therefore $K_{A_0}-K_{A}=\mathcal{E_{\zeta}}$ where $\mathcal{E_{\zeta}}$ is the generator of the null translation on the horizon. Now, we can easily obtain the equation (\ref{D24}) by extracting out the one-sided modular Hamiltonian for the new wedge. Further, one can write the full modular Hamiltonian and use the fact from modular theory that if $\mathcal{M}_{D(A)} \subset \mathcal{M}_{D(A_0)}$  then  $K_{A_0}-K_{A}>0$ for any state. For the class of metrics we consider in this section as well, such a result follows on the Cauchy (Killing) horizon by resumming terms when we only have $\zeta^+(\vec{x})$ deformations or by using HSMI. Further, for this class of metrics, Sewell's result \cite{GLS} for the modular Hamiltonian of the wedge corresponding to the half-space can be used and it is local. This along with the fact that $K_{A_0}-K_{A}>0$  is true for any $\zeta^+>0$ implies,
\begin{equation}\label{D25}
\int_{-\infty}^{\infty}dx_{L+}T_{+ +}(\vec{x}, x_{L+}) \geq 0.
\end{equation}
This is just the Averaged Null energy condition (ANEC) integrated along a null generator of the Cauchy (Killing) horizon. A similar relation can be obtained on $\mathcal{H^-}$. Therefore, in particular, HSMI along with the monotonicity of $K$ implies ANEC along null generators of the future horizon of the Schwarzschild spacetime and for the class of metrics we are working with in this section. This has already been noted by FLPW \cite{TROH}.

\end{document}